%% file: main.tex
\documentclass{article}

 \usepackage[main, final]{neurips_2026}

\usepackage[utf8]{inputenc} 
\usepackage[T1]{fontenc}    
\usepackage{hyperref}       
\usepackage{url}            
\usepackage{booktabs}       
\usepackage{amsfonts}       
\usepackage{nicefrac}       
\usepackage{microtype}      
\usepackage{xcolor}         

\usepackage{graphicx}
\usepackage{subcaption}
\usepackage{booktabs}
\usepackage{multirow}
\usepackage{tabularx}
\usepackage[accsupp]{axessibility}
\usepackage{cleveref}
\usepackage{enumitem}
\usepackage{wrapfig}
\usepackage{titletoc}

\usepackage{makecell}

\title{GLUT: 3D Gaussian Lookup Table for Continuous Color Transformation}

\author{%
  Danna Xue$^{1}$\quad
  David Serrano-Lozano$^{1,2}$\quad
  Shaolin Su$^{1,}$\thanks{Corresponding author.}~\quad
  Javier Vazquez-Corral$^{1,2}$ \\
  $^1${Computer Vision Center} \quad \quad
  $^2${Universitat Autònoma de Barcelona} \\
  \texttt{\{dxue, dserrano, shaolin, javier.vazquez\}@cvc.uab.cat} 
}

\begin{document}

\maketitle

\begin{abstract}
3D Lookup Tables (3D LUTs) are widely used for color mapping, but their grid-based representation discretizes the RGB space, leading to a capacity-memory trade-off that becomes prohibitive when storing large numbers of LUTs. Recent approaches use implicit neural representations to improve scalability, but their black-box nature limits interpretability and localized editing. In this paper, we propose Gaussian LUT (GLUT),  a continuous and explicit representation of color transformations using learnable 3D Gaussian primitives. By eliminating fixed-resolution grids, GLUT provides flexible representational capacity with a compact memory footprint.  Its spatially localized primitives also facilitate interpretation and targeted editing. Building on GLUT, we introduce Conditional Gaussian LUT (CGLUT), a compact generator that predicts GLUT parameters for multiple LUTs within a shared framework, enabling smooth blending between color styles.  GLUT further supports efficient local editing by allowing targeted adjustments to specific color regions without retraining the model. Experiments show that GLUT outperforms existing neural LUT representations in both accuracy and efficiency, while supporting controllable localized color editing. CGLUT further improves compact modeling and smooth blending of multiple color styles. Code is available at \href{https://github.com/CVC-Color/glut}{https://github.com/CVC-Color/glut}.
\end{abstract}

\section{Introduction}
\label{sec:intro}

\begin{figure}[tb]
  \centering
  \begin{subfigure}{0.25\linewidth}
    \centering
    \includegraphics[height=3cm]{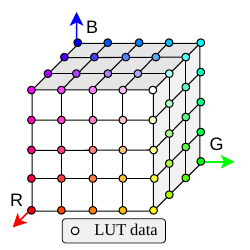}
    \caption{Traditional LUT}
    \label{fig:lut} 
  \end{subfigure}
  \hfill 
  \begin{subfigure}{0.24\linewidth}
    \centering
    \includegraphics[height=3cm]{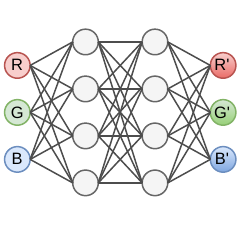}
    \caption{INR-based LUT}
    \label{fig:nilut} 
  \end{subfigure}
  \hfill 
  \begin{subfigure}{0.49\linewidth}
    \centering
    \includegraphics[height=3cm]{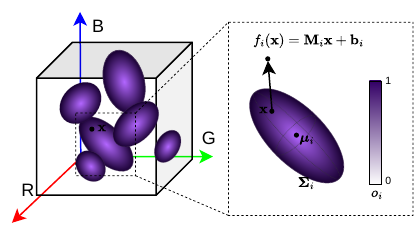}
    \caption{Gaussian LUT (Ours)}
    \label{fig:glut} 
  \end{subfigure}
  \caption{Comparison between different color LUT representations. (a) Explicit Traditional LUT \textit{vs.} (b) Implicit INR-based LUT \textit{vs.} (c) Explicit Gaussian LUT (Ours).}
  \label{fig:lut-comparison}
\end{figure}

3D Lookup Tables (3D LUTs) provide an efficient way to implement deterministic color transformations by storing output colors at sampled points in the color space. They are widely used in real-time image signal processing (ISP) on consumer devices~\cite{karaimer2018improving, kim2023learning, tan2026chameleontuner}, film color grading~\cite{postma2016color, shin2025video, chen2023nlut}, and computational photography~\cite{zhang2022clutnet, yang2022adaint, zeng2020learning, yang2024taming, Kim_2025_ICCV, Yang_2025_CVPR}.
A 3D LUT partitions the RGB space into a regular grid, where each vertex stores a transformed color value, as illustrated in \Cref{fig:lut}. At inference time, output colors are obtained via table lookup followed by interpolation among neighboring grid points. This non-parametric design captures complex nonlinear mappings with high fidelity, and its lookup-based inference offers low computational cost and predictable memory requirements, making 3D LUTs suitable for resource-constrained deployment.

Despite these advantages, traditional 3D LUTs face a fundamental trade-off between representational capacity and memory consumption. Their representational power is tightly coupled to grid resolution: dense sampling enables fine-grained color control but substantially increases storage requirements. In contrast, sparse sampling saves memory but can introduce interpolation artifacts and limit the representation of sharp transitions in color mappings, which may be important for high-contrast cinematic looks and effects such as bleach bypass.

Recent works represent color transformations using implicit neural representations (INRs)~\cite{conde2024nilut, elezabi2024inretouch, zehtab2025efficient}, where a compact network parameterizes a continuous neural LUT (see \Cref{fig:nilut}). While these methods avoid explicit grids and support efficient inference on edge devices, they sacrifice interpretability and local controllability. Since color transformations are encoded across shared network parameters, adjusting a specific color range requires reoptimizing the network weights, limiting practical use in interactive or color-specific editing.

To overcome (1) the capacity-memory trade-off of grid-based LUTs, and (2) the limited interpretability and local controllability in INR-based LUTs, in this paper, we propose a continuous and explicit LUT representation through the lens of explicit volumetric primitives. Inspired by 3D Gaussian Splatting (3DGS)~\cite{kerbl3Dgaussians}, which represents a scene as a set of optimizable, spatially localized Gaussians, we model the color transformation in RGB space using a collection of learnable 3D Gaussian primitives (see \Cref{fig:glut}). 
Specifically, we introduce Gaussian LUT (GLUT), which formulates color transformation as a soft, adaptive partition of the color space using explicit Gaussian anchors. Building on GLUT, we further introduce Conditional Gaussian LUT (CGLUT), a compact generator that predicts GLUT parameters for multiple LUT instances, enabling unified modeling and smooth blending of diverse color styles.

GLUT offers three main advantages. First, its continuous representation decouples representational capacity from a fixed grid resolution, allowing a flexible trade-off between model size and mapping fidelity. Second, its explicit Gaussian primitives provide spatially localized representations of color transformations, facilitating interpretation and targeted editing. Third, localized editing can be performed by optimizing only a subset of relevant primitives, avoiding global model retraining. CGLUT further enables compact modeling and controllable blending of multiple color styles within a shared framework.

Our primary contributions are as follows:
\begin{enumerate}[leftmargin=*,noitemsep]
\item We propose Gaussian LUT (GLUT), an explicit and continuous LUT representation of color transformations based on a compact set of learnable 3D Gaussian primitives.

\item We introduce Conditional Gaussian LUT (CGLUT), a compact generator that predicts the parameters of multiple GLUT instances, enabling unified modeling and smooth, controllable blending of diverse color styles.

\item We develop a localized editing mechanism based on GLUT that adjusts selected Gaussian primitives to modify specific color ranges without retraining the entire model, enabling fine-grained color control with limited interference in unrelated regions.

\item Extensive experiments demonstrate that GLUT achieves favorable accuracy-efficiency trade-offs compared with existing neural LUT representations. Its explicit, spatially localized formulation also supports interpretable color mapping and efficient local editing.
\end{enumerate}


\section{Related Works}
\label{sec:rw}
\noindent \textbf{Color Look-Up Table Representations.} 
3D LUTs provide an efficient mechanism for mapping input colors to desired outputs via a function $(R',G',B')=f(R,G,B)$. A standard 3D LUT discretizes the RGB space into a cubic grid (typically $\text{17}^\text{3}$ or $\text{33}^\text{3}$ vertices), where the outputs are computed through trilinear interpolation or tetrahedral interpolation~\cite{kasson1995performing}. Higher-order schemes such as tricubic splines~\cite{fumoto1997extraction} yield smoother mappings at a higher computational cost.

While computationally efficient, grid LUTs face a trade-off between representational capacity and memory consumption. Existing compression methods reduce storage through hierarchical and differential encoding~\cite{balaji2007hierarchical, shaw2012lossless, tang2016icc}, keypoint-based reconstruction via multi-scale anisotropic diffusion~\cite{tschumperle20193d, tschumperle2020reconstruction}, and low-rank tensor factorization~\cite{monga2012imaging}. While substantially reducing storage, these approaches retain a discrete grid representation or reconstruct one from a more compact encoding, so their spatial resolution and editing behavior remain tied to the underlying grid. Compact parametric alternatives have also been studied for color correction and device characterization. Root-polynomial regression~\cite{finlayson2015color} models color mappings using global polynomial functions, while radial basis function (RBF) interpolation~\cite{poljicak2016optimized} uses localized kernels. These methods are primarily designed for smooth device characterization and color correction, whereas stylized LUTs can contain substantially more complex and localized nonlinear transformations.

More recently, implicit neural representations (INRs) have been introduced for continuous LUT modeling. NILUT~\cite{conde2024nilut} uses a lightweight MLP for efficient color transformation on mobile deployment and supports LUT blending through embedding interpolation. INRetouch~\cite{elezabi2024inretouch} incorporates convolutional features for content-aware local color transformation, while ENNELUT~\cite{zehtab2025efficient} uses an invertible network to represent hundreds of LUTs within a single model. However, neural implicit functions exhibit a spectral bias toward smooth, low-frequency mappings~\cite{rahaman2019spectral,tancik2020fourier}, which can make sharp local transitions difficult to represent accurately. Moreover, their fully implicit parameterization provides no explicit spatial decomposition of the RGB space, making localized modification less direct. In contrast, GLUT represents color transformations using spatially localized Gaussian primitives, providing an explicit structure for color mapping and local editing.

\noindent \textbf{Gaussian-based Representations.} 
Representing a function as a Gaussian-weighted combination of local models has a long history. Radial basis function (RBF) networks~\cite{moody1989fast} approximate functions with localized basis functions, and RBF interpolation has been applied to color characterization~\cite{poljicak2016optimized}. Mixtures of local experts~\cite{jacobs1991adaptive} combine specialized models through input-dependent gating, while GLLiM~\cite{deleforge2015high} couples Gaussian gating with locally affine regressors and fits its parameters through expectation-maximization (EM). These methods demonstrate the effectiveness of localized function approximation, but differ in their parameterizations and optimization objectives. They do not specifically target a compact, continuous, and directly editable representation of complex color transformations.

More recently, Gaussian Splatting~\cite{kerbl3Dgaussians, huang20242d} has emerged as a powerful alternative to implicit neural fields like NeRF~\cite{mildenhall2021nerf}, offering superior rendering fidelity and speed. By representing complex 3D scenes with optimizable Gaussian primitives, these methods enable efficient rendering and direct manipulation of explicit scene elements. This paradigm has also been extended beyond 3D reconstruction and editing~\cite{kerbl3Dgaussians, huang20242d, wu2024gaussctrl, barda2025instant3dit} to low-level vision tasks such as image restoration~\cite{chen2024deblur}, image compression~\cite{zhang2024gaussianimage, zhang2025image}, and super-resolution~\cite{chen2025generalized, yang2024mob}. 

Inspired by these explicit, localized representations, we transfer the Gaussian-primitive paradigm from the spatial domain to the RGB color space. GLUT represents a color transformation as a weighted combination of Gaussian-localized affine transformations, providing a continuous representation with explicit local structure that supports targeted color editing.

\noindent \textbf{Color Manipulation with LUTs.}
Conventional LUTs apply a global color mapping independently of image content. To support context-dependent transformations, prior work has explored image-adaptive LUTs by combining multiple basis LUTs~\cite{zeng2020learning}. Subsequent methods improve adaptation through adaptive sampling~\cite{yang2022adaint}, context-aware weighting~\cite{liu20234d}, spatially aware bilateral grids~\cite{kim2024image}, and text-guided control~\cite{lee2024cliptone}. Other approaches focus on accelerating LUT-based processing~\cite{yang2024taming, yang2022seplut}, or generating LUTs directly using diffusion models or vision-language models (VLMs)~\cite{shin2025video, ma2026acetone}. These methods primarily address how LUTs are predicted, conditioned, or adapted to image content. Conversely, our work focuses on the LUT representation itself. Given a target color transformation, GLUT seeks to encode its mapping in a compact, continuous, and explicitly structured form while supporting localized editing.


\section{Method}
\label{sec:method}

We propose 3D Gaussian LUT (GLUT), an explicit representation for continuous color transformations using a compact set of learnable 3D Gaussian primitives. \Cref{sec:GLUT} introduces the GLUT formulation. \Cref{sec:CGLUT} presents a conditional generator that predicts GLUT parameters for multiple LUT styles within a shared framework. Finally, in \Cref{sec:edit}, we present an efficient LUT editing mechanism enabled by the localized structure of Gaussian primitives.

\subsection{3D Gaussian LUT}
\label{sec:GLUT}

GLUT represents a color mapping using a set of learnable 3D Gaussian primitives in RGB space, as illustrated in \Cref{fig:glut}. Unlike INR-based LUTs, which encode color transformations in shared network parameters, GLUT explicitly associates each primitive with a spatially localized affine transformation.

\noindent \textbf{Model Parameters.}
The representation consists of $N$ 3D Gaussian primitives, where each primitive $i$ is defined by:
\begin{itemize}[leftmargin=2em,noitemsep]
    \item \textit{Mean} $\boldsymbol{\mu}_i \in \mathbb{R}^3$, representing its center in RGB space.
    \item \textit{Covariance matrix} $\boldsymbol{\Sigma}_i \in \mathbb{R}^{3 \times 3}$, controlling spatial extent and orientation. To ensure positive definiteness of each covariance matrix, we use the Cholesky parameterization $\boldsymbol{\Sigma}_i = \mathbf{L}_i \mathbf{L}_i^\top$, where $\mathbf{L}_i$ is a lower-triangular matrix whose diagonal entries are constrained to be positive using the Softplus function. We parameterize each covariance matrix using six learnable parameters per Gaussian.
    \item \textit{Opacity} $o_i \in [0,1]$, determining the contribution weight of the Gaussian.
    \item \textit{Local color transformation}: parameterized by a transformation matrix $\mathbf{M}_i \in \mathbb{R}^{3 \times 3}$ and a bias term $\mathbf{b}_i \in \mathbb{R}^3$. 
\end{itemize}

In addition, we introduce a \textit{global base transformation} for each GLUT, parameterized by a matrix $\mathbf{G} \in \mathbb{R}^{3 \times 3}$ and bias $\mathbf{g} \in \mathbb{R}^3$, which captures coarse global color shifts. All the GLUT parameters $\mathbf{p}=\{\boldsymbol{\mu}_i, \boldsymbol{\Sigma}_i, o_{i}, \mathbf{M}_{i}, \mathbf{b}_{i}\}_{i=1}^{N}\bigcup \{ \mathbf{G}, \mathbf{g} \}$ are learnable.

\vspace{3mm}
\noindent \textbf{Color Transformation via GLUT.}
Given an input color $\mathbf{x} \in [0,1]^3$, we compute the continuous color mapping $f(\mathbf{x})$ by combining the global transformation with a set of weighted, localized Gaussian transformations. 

\noindent \textit{\textbf{Gaussian weighting}.}
The influence of each Gaussian transformation is defined by the density of its corresponding Gaussian primitive in RGB space:
\begin{equation}
p_i(\mathbf{x}) 
=
\frac{1}{\sqrt{(2\pi)^{3} |\boldsymbol{\Sigma}_i|}}
\exp\!\left(
-\frac{1}{2}
d_{i}(\mathbf{x})^2
\right),
\end{equation}

\noindent where $d_{i}(\mathbf{x})=\sqrt{(\mathbf{x}-\boldsymbol{\mu}_i)^\top
\boldsymbol{\Sigma}_i^{-1}(\mathbf{x}-\boldsymbol{\mu}_i)}$ denotes the Mahalanobis distance from $\mathbf{x}$ to Gaussian $i$.

\noindent Combining density with opacity $o_i$, we obtain a normalized influence weight of Gaussian $i$:
\begin{equation}
w_i(\mathbf{x})
=
\frac{p_i(\mathbf{x})\, o_i}
{\sum_{j=1}^{N} p_j(\mathbf{x})\, o_j + \epsilon},
\label{eq:weights}
\end{equation}

\noindent where $\epsilon$ is a small constant to ensure numerical stability. This formulation yields a soft partition of the RGB space.

\noindent \textit{\textbf{Final color output.}}
Each Gaussian applies a local affine transformation:
\begin{equation}
f_i(\mathbf{x})=\mathbf{M}_i \mathbf{x}+\mathbf{b}_i.
\end{equation}

\noindent In addition, the global affine transform provides a coarse base mapping: 
\begin{equation}
f_{\text{global}}(\mathbf{x})=\mathbf{G}\mathbf{x}+\mathbf{g}.
\end{equation}

\noindent The final color output is obtained via a mixture of all local and global transformations:
\begin{equation}
f(\mathbf{x})=\sum_{i=1}^{N} w_i(\mathbf{x})f_i(\mathbf{x})+f_{\text{global}}(\mathbf{x}).
\label{eq:mapping}
\end{equation}

\noindent The value of $\mathbf{\hat{y}}=f(\mathbf{x})$ is clamped to $[0,1]^3$ to ensure valid RGB values.

\vspace{3mm}
\noindent \textbf{Training Objective.}
\label{sec:glut_loss}
Since GLUT is an explicit representation, all the parameters are learnable and can be directly optimized using a loss function consisting of three terms:

\noindent \textbf{\textit{Reconstruction loss.}}
We use the $\ell_1$ reconstruction loss between the predicted output $\mathbf{\hat{y}} = f(\mathbf{x})$ and the target color $\mathbf{y}$:
\begin{equation}
\mathcal{L}_{\text{rec}}=\left\|\mathbf{\hat{y}}- \mathbf{y}\right\|_1.
\end{equation}

\noindent \textbf{\textit{Hue-Chroma loss.}}
We further introduce a hue-chroma loss in the perceptually uniform CIELAB color space. Given the predicted color $\mathbf{\hat{y}}$ and the target color $\mathbf{y}$, we first transform both into the CIELAB space. For a color with CIELAB components $(L^,a^,b^)$, its chroma is defined as $C = \sqrt{a^2+b^2}$ and the hue is represented as a 2D unit vector $\mathbf{h}=(\frac{a}{C}, \frac{b}{C})$. We define the loss as the cosine distance between the predicted and target hue vectors, weighted by the target chroma:
\begin{equation}
\mathcal{L}_{\text{hc}} = C \cdot (1 - \langle \mathbf{\hat{h}}, \mathbf{h} \rangle).
\end{equation}
\noindent This formulation prioritizes hue accuracy in highly saturated regions while maintaining tolerance in low-chroma regions (\eg, neutrals), where hue is less perceptually meaningful and more susceptible to noise.

\noindent \textbf{\textit{Sparsity regularization.}}
To encourage specialization of Gaussian primitives, we regularize their opacity parameters using a binary entropy term:
\begin{equation}
\label{eq:sparsity}
\mathcal{R}_{\text{sparse}}=-\frac{1}{N}
\sum_{i=1}^{N}\left[o_i \log(o_i + \epsilon)+(1 - o_i) \log(1 - o_i + \epsilon)\right],
\end{equation}
\noindent where $\epsilon$ is a small constant for numerical stability.

\noindent \textbf{\textit{Total loss.}} The overall training objective is $\mathcal{L}_{\text{total}}=
\mathcal{L}_{\text{rec}}+\lambda_{\text{hc}} \mathcal{L}_{\text{hc}}+
\lambda_{\text{sparse}} \mathcal{R}_{\text{sparse}}$,
where $\lambda_{\text{hc}}$ and $\lambda_{\text{sparse}}$ control the contributions of the hue-chroma loss and sparsity regularization, respectively.

\noindent In summary, GLUT represents a color transformation as a continuous mapping that does not depend on a fixed sampling grid. It can model complex nonlinear transformations with fewer parameters than implicit neural networks. Each Gaussian primarily governs a local region of the color space, which enables targeted editing. The Gaussian parameters also have clear geometric meanings: the mean identifies the center of the affected color region, and the covariance describes the shape and extent of the region of influence. The entire pipeline is differentiable, so the GLUT parameters $\mathbf{p}$ can be optimized end-to-end, either from color tables or directly from natural images.

\subsection{Conditional Gaussian LUT}
\label{sec:CGLUT}

Building upon the single Gaussian LUT formulation, we extend the model to represent multiple color transformations within a unified conditional framework, termed Conditional Gaussian LUT (CGLUT). Instead of representing each LUT independently using a one-hot condition vector, as in~\cite{conde2024nilut, zehtab2025efficient}, we condition the transformation on a learnable embedding that encodes each color style (see \Cref{fig:cglut}). This enables a single model to represent multiple transformation functions while sharing structural priors across conditions.

\textbf{\begin{figure}[t]
  \centering
  \includegraphics[width=\linewidth]{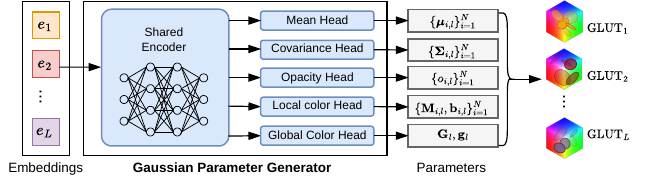}
  \caption{Structure of Conditional Gaussian LUT (CGLUT). The Gaussian parameter generator maps the learned embedding to the parameters of all Gaussian primitives in a GLUT using a shared MLP encoder and parameter-specific prediction heads.}
  \label{fig:cglut}
\end{figure}}

\noindent \textbf{Condition Embedding.}
Let $l \in \{1,\ldots,L\}$ denote the condition index of the $l$th LUT. We introduce a learnable embedding matrix $\mathbf{E} \in \mathbb{R}^{L \times D}$, where $D$ is the embedding dimension. For a given condition $l$, the embedding vector $\mathbf{e}_l\in \mathbb{R}^D$ serves as a compact latent representation of the corresponding color style.

\noindent \textbf{Gaussian Parameter Generation.}
We adopt a Gaussian parameter generator $\mathcal{G}$ that maps the embedding $\mathbf{e}_l$ to the parameters of all Gaussian primitives in a Gaussian LUT: $\mathbf{p}_l = \mathcal{G}(\mathbf{e}_l)$. The generator consists of a shared encoder that extracts common features, followed by multiple prediction heads, where each head predicts one type of parameter of the Gaussians. The generator is implemented using MLPs. Given a condition index $l$, the generator first predicts the corresponding GLUT parameters $\mathbf{p}_l = \mathcal{G}(\mathbf{e}_l)$. The input color $\mathbf{x}$ is then transformed using the predicted parameters, following the same GLUT formulation as in the single-LUT case. The training objectives are the same as those of GLUT (\Cref{sec:GLUT}), while the learnable parameters in CGLUT comprise the embeddings $\mathbf{e}_l$ and the parameters of the generator $\mathcal{G}$.

\noindent \textbf{LUT Blending.}
Since each LUT is represented by a continuous embedding vector in CGLUT, we blend two LUTs, indexed by $l_1$ and $l_2$, by linearly interpolating their embeddings. The interpolated embedding $\mathbf{e}_{l_1l_2}^{\alpha} = (1-\alpha)\mathbf{e}_{l_1} + \alpha \mathbf{e}_{l_2}$, $\alpha \in [0,1]$ is fed into the generator to obtain the corresponding Gaussian parameters, yielding smooth transitions between color transformations. 

In addition to generating all GLUT parameters independently for each condition (\textit{Full Generation}), we also consider an alternative configuration, termed \textit{Shared Geometry}. In this setup, the spatial attributes, Gaussian positions and shapes $\{\boldsymbol{\mu}_{i}, \boldsymbol{\Sigma}_{i}\}_{i=1}^{N}$, are shared across all styles, while the opacities and global color transformation parameters $\{o_{i,l}, \mathbf{M}_{i,l}, \mathbf{b}_{i,l}, \mathbf{G}_l, \mathbf{g}_l\}_{i=1}^{N}$ are generated conditionally. While the \textit{Full Generation} setup offers greater flexibility for better reconstruction quality, \textit{Shared Geometry} facilitates smoother color style transitions through embedding interpolation, as it maintains a consistent spatial partition of the color manifold.

Overall, CGLUT preserves the continuous and localized nature of GLUT, while enabling compact multi-style modeling with a single network.

\subsection{GLUT Editing}
\label{sec:edit}
A key advantage of GLUT over INR-based LUTs~\cite{conde2024nilut, zehtab2025efficient} is that its parameters carry explicit geometric interpretations in RGB space, enabling direct and interpretable editing without retraining. Although users could manipulate Gaussian parameters directly, we found this approach unintuitive in practice, as multiple primitives must be adjusted simultaneously to achieve a coherent result. Instead, we propose a user-friendly editing interface in which the user specifies a \emph{color constraint} $\mathbf{c} = (\mathbf{c}_\text{in}, \mathbf{c}_\text{out})$, where $\mathbf{c}_\text{in} \in [0,1]^3$ is an input color to modify and $\mathbf{c}_\text{out} \in [0,1]^3$ is the desired output color. The goal is to update a minimal subset of GLUT parameters such that $f(\mathbf{c}_\text{in}) \approx \mathbf{c}_\text{out}$, while minimizing unintended changes elsewhere in the color space. This formulation naturally supports iterative, fine-grained adjustments, allowing users to steer the learned color transformation towards a desired style without global interference or model retraining.

Given the current GLUT output at $\mathbf{c}_\text{in}$, we compute the correction residual:
\begin{equation}
    \boldsymbol{\delta} = \mathbf{c}_\text{out} - f(\mathbf{c}_\text{in})
    \in \mathbb{R}^3.
    \label{eq:residual}
\end{equation}
Rather than modifying all $N$ Gaussians, we select the $K$ most influential ones at $\mathbf{c}_\text{in}$ based on the normalized influence weights from Eq.~\eqref{eq:weights}: 
\begin{equation}
    \mathcal{K} = \operatorname{argtop}_K \bigl\{ w_i(\mathbf{c}_\text{in}) \bigr\}_{i=1}^N.
    \label{eq:topk}
\end{equation}
Although $\mathcal{K}$ can be chosen freely by the user, selecting the $K$ most influential Gaussians at $\mathbf{c}_\text{in}$ provides a natural and automatic proxy. The parameter $K$ controls the spatial extent of the edit: a smaller $K$ confines the modification to the immediate neighbourhood of $\mathbf{c}_\text{in}$, whereas a larger $K$ distributes $\boldsymbol{\delta}$ across more Gaussians, producing a broader effect in the color space.

We distribute the residual $\boldsymbol{\delta}$ among the selected Gaussians proportionally to their normalised influence:
\begin{equation}
    \alpha_k = \frac{w_k(\mathbf{c}_\text{in})}{\sum_{j \in \mathcal{K}}
    w_j(\mathbf{c}_\text{in})}, \quad k \in \mathcal{K}.
    \label{eq:alpha}
\end{equation}
The bias term of each selected Gaussian is then updated as:
\begin{equation}
    \mathbf{b}_k \leftarrow \mathbf{b}_k + s \cdot \alpha_k \cdot \boldsymbol{\delta},
    \quad k \in \mathcal{K},
    \label{eq:bias_update}
\end{equation}
where $s \in [0,1]$ is a user-controlled \emph{strength} parameter. We update only the bias terms $\mathbf{b}_k$ and keep the local color transformation matrices $\mathbf{M}_k$ fixed. This better preserves the local color relationships (\ie contrast, saturation, and structure) encoded by each Gaussian, restricting the edit to a global shift of the output color within that region.

The edit is performed in real time and is inherently localized, without requiring an explicit locality constraint. The densities $p_k$ of the modified Gaussians decay exponentially with the Mahalanobis distance from their centers. Consequently, at colors $\mathbf{c}$ where other primitives dominate, the normalized weights of the modified Gaussians tend to be small, making the effect of the bias update on $f(\mathbf{c})$ negligible. This locality naturally arises from the explicit Gaussian representation in color space. An interactive GLUT editing tool is provided in the Appendix, with further details in~\Cref{sup_sec:editing}.


\section{Experiments}
\label{sec:exp}

\subsection{Experimental Setup}
\label{sec:exp_setup}

\noindent \textbf{Implementation Details.}
The model is implemented in PyTorch and optimized using the Adam optimizer on a single NVIDIA GeForce RTX 4090 GPU. We employ a cosine annealing learning rate schedule, starting from $10^{-3}$ over the entire training duration. GLUT is trained for 20 epochs with a batch size of 1024, while CGLUT is trained for 40 epochs with a batch size of 8192. The loss weights are empirically set to $\lambda_\text{hc}=10$ and $\lambda_\text{sparse}=0.001$. Additional implementation details are provided in~\Cref{sup_sec:imple}.

\noindent \textbf{Dataset.}
Our dataset comprises 300 LUTs collected from publicly available online sources, split into 225 LUTs with resolutions of $\text{33}^\text{3}$ or $\text{32}^\text{3}$ and 75 LUTs at $\text{64}^\text{3}$. The 7 LUTs used for multi-LUT modeling are taken from \cite{conde2024nilut}, each defined on a standard $\text{33}^\text{3}$ grid, and are also included in our 225-LUT subset. For training, we extract dense color mappings from the corresponding \texttt{.cube} files. We uniformly sample the full 8-bit RGB space to construct a $\text{128}^\text{3}$ training sample, reserving the remaining colors for evaluation to verify that the model learns a continuous representation. The resulting mappings are reshaped into Hald images \cite{haldwebsite} of resolutions 1024$\times$2048$\times$3 and 3584$\times$4096$\times$3 for the training and test colors, respectively. All LUTs are open-source under Creative Commons licensing.

\noindent \textbf{Evaluation.}
Color fidelity is measured via PSNR, CIEDE2000 ($\Delta E_{00}$), and CIEDE1976 ($\Delta E_{76}$), covering both RGB and perceptual CIELAB color spaces. We evaluate GLUT on two settings: direct Hald image retouching and applying learned neural LUTs to 100 natural images from MIT-Adobe FiveK~\cite{bychkovsky2011learning}, where we additionally report SSIM~\cite{wang2004image} and LPIPS~\cite{zhang2018unreasonable} for structural similarity and perceptual quality. 
Model efficiency is reported in GFLOPs (measured on 512$\times$512$\times$3 image) and parameter count. \textbf{Bold} and \textit{italic} indicate the best and second-best results in each table, respectively.

\noindent \textbf{Comparison Methods.}
We evaluate GLUT against NILUT~\cite{conde2024nilut} on single-LUT modeling, and CGLUT against CNILUT~\cite{conde2024nilut} and ENNELUT~\cite{zehtab2025efficient} on multi-LUT representation. NILUT and CNILUT follow official configurations with extended training iterations (10K and 60K, respectively) to ensure adequate convergence. ENNELUT is re-implemented because its code is unavailable. NILUT (64$\times$2) uses two hidden layers of 64 neurons each, while GLUT-16 uses 16 Gaussian primitives. ENNELUT is trained with $L_2$ loss or $\Delta E$ loss, respectively. INRetouch~\cite{elezabi2024inretouch} addresses local retouching rather than global color transformations, so it is compared separately in~\Cref{sup_sec:inretouch} in the Appendix.

\subsection{Experiment Results}

\noindent \textbf{Comparison with classical approaches.}
We compare GLUT with classical approaches from three categories: \textit{root-polynomial regression}~\cite{finlayson2015color}, \textit{Gaussian-based local models}: RBF network~\cite{moody1989fast} and Gaussian Locally Linear Mapping (GLLiM)~\cite{deleforge2015high}, and \textit{LUT compression and interpolation}: Keypoint diffusion~\cite{tschumperle2020reconstruction}, Tetrahedral
interpolation~\cite{kasson1995performing}, and Tricubic interpolation~\cite{fumoto1997extraction}. For all baselines except root-polynomial regression~\cite{finlayson2015color}, we match the parameter count to that of GLUT-32 for a controlled comparison. The implementation details are provided in \Cref{sup_sec:classic} in the Appendix.
As shown in Table~\ref{tab:classic}, GLUT substantially outperforms the RBF network, demonstrating the benefit of modeling each Gaussian primitive with a local affine transformation rather than a constant output. The performance gap between the two GLLiM variants further suggests that our gradient-descent (GD) optimization framework provides a more effective fit than classical expectation-maximization (EM). The relative ranking of the methods remains consistent across the 75-LUT and 225-LUT datasets, with GLUT-32 achieving the best PSNR and $\Delta E_{00}$ under a matched parameter budget in both settings.

\begin{table}[t]
\centering
\caption{Quantitative comparison between classic baselines and GLUT.}
\label{tab:classic}
\setlength{\tabcolsep}{4pt}
\resizebox{0.7\textwidth}{!}{%
\begin{tabular}{l|ccc|ccc|c}
\toprule
\multicolumn{1}{c|}{\multirow{2}{*}{Method}} & \multicolumn{3}{c|}{75 LUTs} & \multicolumn{3}{c|}{225 LUTs} & \multirow{2}{*}{\#Params} \\
\multicolumn{1}{c|}{} & PSNR$\uparrow$ & $\Delta E_{00}$$\downarrow$ & $\Delta E_{76}$$\downarrow$ & PSNR$\uparrow$ & $\Delta E_{00}$$\downarrow$ & $\Delta E_{76}$$\downarrow$ &  \\
\midrule
Root-polynomial regression \cite{finlayson2015color} & 23.55 & 5.10 & 9.96 & 25.15 & 4.66 & 9.23 & 18 \\ \midrule
RBF network (Our GD) \cite{moody1989fast} & 42.90 & 0.52 & 0.96 & 47.82 & 0.38 & 0.73 & 717 \\
GLLiM (EM) \cite{deleforge2015high} & 38.03 & 1.01 & 1.93 & 41.88 & 0.78 & 1.50 & 726 \\
GLLiM (Our GD) \cite{deleforge2015high} & \textit{44.72} & \textit{0.44} & \textit{0.82} & \textit{49.13} & \textit{0.34} & \textit{0.66} & 726 \\ \midrule
Keypoint diffusion \cite{tschumperle2020reconstruction} & 31.41 & 1.75 & 3.64 & 33.81 & 1.82 & 3.96 & 720 \\
$6^3$ LUT, tetrahedral \cite{kasson1995performing} & 33.55 & 1.47 & 3.20 & 36.20 & 1.31 & 2.87 & 648 \\
$6^3$ LUT, tricubic \cite{fumoto1997extraction} & 34.81 & 1.22 & 2.49 & 38.51 & 1.04 & 2.12 & 648 \\ \midrule
GLUT-32 (Ours) & \textbf{45.47} & \textbf{0.41} & \textbf{0.77} & \textbf{50.42} & \textbf{0.30} & \textbf{0.58} & 716 \\ \bottomrule
\end{tabular}
}
\end{table}

\begin{table}[t]
\centering
\caption{Quantitative comparison of single-LUT modeling on Hald images and natural images.}
\label{tab:glut-hald}
\resizebox{\columnwidth}{!}{%
\begin{tabular}{l|ccc|ccccc|c|c}
\toprule
\multicolumn{1}{c|}{\multirow{2}{*}{Method}} & \multicolumn{3}{c|}{Hald images} & \multicolumn{5}{c|}{Natural images} & \multirow{2}{*}{\#Params} & \multirow{2}{*}{GFLOPs} \\
\multicolumn{1}{c|}{} & PSNR$\uparrow$ & $\Delta E_{00}\downarrow$ & $\Delta E_{76}\downarrow$ & PSNR$\uparrow$ & SSIM$\uparrow$ & LPIPS$\downarrow$ & $\Delta E_{00}\downarrow$ & $\Delta E_{76}\downarrow$ &  &  \\ \midrule
NILUT (64$\times$2)~\cite{conde2024nilut} & 36.37 & 1.18 & 2.26 & 38.17 & 0.979 & 0.017 & 2.95 & 3.82 & 8,771 & 4.60 \\
NILUT (128$\times$2)~\cite{conde2024nilut} & 38.84 & 0.86 & 1.62 & 40.84 & 0.990 & 0.009 & 2.07 & 2.54 & 33,923 & 17.78 \\
GLUT-16 & 41.50 & 0.64 & 1.22 & 43.01 & 0.994 & 0.005 & 1.31 & 1.75 & 364 & 0.25 \\
GLUT-32 & \textit{45.47} & \textit{0.41} & \textit{0.77} & \textit{45.92} & \textit{0.997} & \textit{0.003} & \textit{0.98} & \textit{1.25} & 716 & 0.49 \\
GLUT-64 & \textbf{48.42} & \textbf{0.31} & \textbf{0.56} & \textbf{47.90} & \textbf{0.998} & \textbf{0.002} & \textbf{0.76} & \textbf{0.91} & 1,420 & 0.98 \\\bottomrule
\end{tabular}%
}
\end{table}

\noindent \textbf{Single GLUT.}
\Cref{tab:glut-hald} reports the quantitative results of GLUT and NILUT on 75-LUT Hald images and 100 natural images. GLUT achieves higher PSNR and lower $\Delta E$ with substantially fewer parameters and less computation. For a GLUT model with $N$ Gaussian primitives, the total parameter count is $22N + 12$. In a typical configuration with $N=32$, the model consists of only 716 parameters, resulting in a significantly smaller memory footprint than INR-based models. Results on 225-LUT Hald images and additional qualitative comparisons are shown in \Cref{sup_tab:glut-hald} and \Cref{fig:glut-image1} of \Cref{sup_sec:glut} in the Appendix.

\begin{table}[t]
\caption{Quantitative comparison of multiple-LUT modeling on Hald image dataset.}
\label{tab:cglut}
\resizebox{\linewidth}{!}{%
\setlength{\tabcolsep}{1.5pt}
\begin{tabular}{l|ccccc|ccccc|ccccc}
\toprule
\multicolumn{1}{c|}{\multirow{2}{*}{Method}} & \multicolumn{5}{c|}{7 LUTs}                                   & \multicolumn{5}{c|}{75 LUTs}                                  & \multicolumn{5}{c}{225 LUTs}                                 \\ 
\multicolumn{1}{c|}{}                        & PSNR$\uparrow$  & $\Delta E_{00}\downarrow$ & $\Delta E_{76}\downarrow$ & \#Params  & GFLOPs & PSNR$\uparrow$  & $\Delta E_{00}\downarrow$ & $\Delta E_{76}\downarrow$ & \#Params  & GFLOPs & PSNR$\uparrow$  & $\Delta E_{00}\downarrow$ & $\Delta E_{76}\downarrow$ & \#Params  & GFLOPs  \\ \midrule
CNILUT (128$\times$3)~\cite{conde2024nilut}                              & 44.90 & 0.43            & 0.85            & 51K  & 26.92  & 38.60 & 0.99            & 1.88            & 60K  & 31.48  & 38.52 & 1.16            & 2.21            & 79K  & 41.54  \\
CNILUT (256$\times$3)~\cite{conde2024nilut}                              & 48.38 & 0.32            & 0.61            & 201K & 105.36 & 41.20 & 0.72            & 1.33            & 218K & 114.48 & 41.46 & 0.80            & 1.52            & 257K & 134.62 \\ \midrule
ENNELUT ($\Delta E$)~\cite{zehtab2025efficient}                                & 49.99 & 0.27            & 0.55            & 14K  & 3.55   & 41.82 & 0.62            & 1.24            & 20K  & 5.26   & 39.43 & 0.95            & 1.95            & 34K  & 9.03   \\
ENNELUT ($L_2$)~\cite{zehtab2025efficient}                                 & 53.07 & 0.20            & 0.40            & 14K  & 3.55   & 44.29 & 0.55            & 1.00            & 20K  & 5.26   & 40.68 & 0.96            & 1.87            & 34K  & 9.03   \\
 \midrule
CGLUT-32 (Small)                             & 50.76 & 0.21           & 0.41           & 84K  & 0.49   & 45.66 & 0.39           & 0.73           & 89K & 0.49   & 46.68 & 0.41           & 0.82           & 98K  & 0.49   \\
CGLUT-32 (Large)                            & 53.55 & 0.17            & 0.34            & 233K & 0.49   & 46.69 & 0.36            & 0.66            & 238K & 0.49   & \textit{49.84} & \textit{0.30 }           & \textit{0.58}            & 247K & 0.49   \\
CGLUT-64 (Small)                             & \textit{54.06} & \textit{0.16}           & \textit{0.32}           & 130K & 0.98   & \textit{48.43} & \textit{0.30}           & \textit{0.55}           & 135K & 0.98   & 49.39 & \textit{0.30}           & 0.59           & 144K & 0.98  \\ 
CGLUT-64 (Large)                             & \textbf{55.10} & \textbf{0.14}            & \textbf{0.29}            & 324K & 0.98   & \textbf{49.37} & \textbf{0.28}            & \textbf{0.51}            & 328K & 0.98   & \textbf{52.41} & \textbf{0.22}            & \textbf{0.43}            & 338K & 0.98    \\ 

\bottomrule
\end{tabular}%
}
\end{table}

\noindent \textbf{Multiple LUTs.} 
\Cref{tab:cglut} and \Cref{sup_tab:cglut_image} report the performance of CGLUT in representing multiple color styles. We evaluate CGLUT on 7, 75, and 225 LUTs, comparing it against CNILUT across two model scales. 
We consider two CGLUT variants, Small and Large, with increasing modeling capacity by using 64 and 128 neurons, respectively, in each layer of the parameter generator.

During inference, CGLUT maintains a constant input dimensionality regardless of the library size, \ie, the number of learned LUTs. The parameter generation process is executed only once per style, incurring negligible computational cost (about 4 MFLOPs for CGLUT-Large). The generated parameters are then used for pixel-wise color mapping. Consequently, the computational cost of CGLUT lies in the color transformation stage rather than the conditioning mechanism. This decoupled architecture keeps the inference cost on a natural image largely independent of the number of LUTs in the library. In contrast, CNILUT's computational cost scales with the library size because the style condition must be processed at each pixel, resulting in substantial overhead for large LUT collections.

In terms of reconstruction quality, CGLUT outperforms both CNILUT and ENNELUT on the 75- and 225-LUT datasets. In the 7-LUT setting, only ENNELUT ($L_2$) achieves better results than CGLUT-32 (Small), further demonstrating the competitive performance of our explicit framework.

\noindent \textbf{Efficiency and Model Size.}
\Cref{tab:runtime} summarizes the runtime performance on an AMD EPYC 9634 CPU (84-core) and an NVIDIA RTX 4090 GPU (24 GB), along with the compression ratios of different neural LUTs. Runtime is measured as the average processing time over 100 natural images from MIT5K \cite{bychkovsky2011learning} at three resolutions: Small (512$\times$512), HD (1280$\times$720), and 4K UHD (3840$\times$2160). 
The compression ratio is defined as the size of the neural LUT (\texttt{.pth} file) relative to that of the standard grid-based LUT (\texttt{.cube} file).

Grid-based LUTs remain the fastest option for applying a fixed color transformation, as GPU texture units perform lookup and interpolation in hardware. Among compact representations, GLUT achieves the highest efficiency. For single-LUT models, GLUT-16 runs at approximately 1 FPS on a 4K image on the CPU and at 241 FPS with our fused CUDA kernel, compared with 47 FPS for NILUT (64$\times$2) accelerated with tiny-cuda-nn. For multi-LUT models, CGLUT is the second fastest on the CPU, after ENNELUT, and it reaches 194 FPS at 4K with CUDA, compared with 10 FPS for ENNELUT. At high resolutions, the cost of CGLUT is dominated by applying the generated GLUT, since the generator runs once per image whereas the color transform runs once per pixel. Accordingly, the Small and Large variants achieve nearly identical speeds of 194 and 193 FPS on 4K images.

\begin{wraptable}{r}{0.64\textwidth}
\centering
\caption{Comparison of runtime and compression ratio. ``-'' denotes FPS$<$0.01. ``$\dag$'' denotes time on GPU texture unit. ``$\star$'' denotes implementation with tiny-cuda-nn~\cite{muller2022instant}.}
\label{tab:runtime}
\setlength{\tabcolsep}{3pt}
\resizebox{0.63\columnwidth}{!}{%
\begin{tabular}{l|ccc|ccc|cc}
\toprule
\multicolumn{1}{c|}{\multirow{2}{*}{Method}} & \multicolumn{3}{c|}{FPS (CPU)$\uparrow$} & \multicolumn{3}{c|}{FPS (GPU)$\uparrow$} & \multicolumn{2}{c}{Comp. Ratio (\%)$\downarrow$} \\ 
\multicolumn{1}{c|}{} & $\text{512}^\text{2}$ & HD & 4K & $\text{512}^\text{2}$ & HD & 4K & 75-LUT & 225-LUT \\ \midrule
Grid LUT & 3489 & 759 & 148 & 26673$^\dag$ & 5577$^\dag$ & 1510$^\dag$ & 100 & 100 \\ \midrule
NILUT (64$\times$2)~\cite{conde2024nilut} & 11.42 & 1.82 & 0.47 & 1538$^\star$ & 239$^\star$ & 47$^\star$ & 0.55 & 4.32 \\
NILUT (128$\times$2)~\cite{conde2024nilut} & 5.87 & 0.83 & 0.20 & 822$^\star$ & 111$^\star$ & 28$^\star$ & 1.96 & 15.70 \\ 
GLUT-16 & 30.12 & 3.69 & 0.97 & 6496 & 952 & 241 & 0.08 & 0.60 \\
GLUT-32 & 14.05 & 1.96 & 0.49 & 6170 & 891 & 226 & 0.10 & 0.75 \\
GLUT-64 & 6.80 & 1.02 & 0.25 & 4917 & 693 & 176 & 0.14 & 1.05 \\\midrule \midrule
CNILUT (128$\times$3)~\cite{conde2024nilut} & 0.34 & 0.01 & - & 1.86 & 0.03  & - & 0.05 & 0.16 \\
CNILUT (256$\times$3)~\cite{conde2024nilut} & 0.30 & - & - & 1.83 & 0.03 & - & 0.17 & 0.52 \\
ENNELUT ($L_2$)~\cite{zehtab2025efficient} & 4.92 & 0.66 & 0.18 & 406 & 39 & 10 & 0.02 & 0.07 \\
CGLUT-32 (Small) & 1.01 & 0.14 & 0.04 & 1147 & 545 & 194 & 0.07 & 0.20 \\
CGLUT-32 (Large) & 0.87 & 0.13 & 0.03 & 1146 & 545 & 193 & 0.18 & 0.50 \\
CGLUT-64 (Small) & 0.52 & 0.07 & 0.02 & 1100 & 473 & 160 & 0.10 & 0.30 \\ 
CGLUT-64 (Large) & 0.48 & 0.07 & 0.02 & 1100 & 465 & 156 & 0.25 & 0.68 \\\bottomrule

\end{tabular}}
\end{wraptable}

Furthermore, neural LUTs offer a compact, continuous representation that significantly reduces memory overhead. For instance, standard $\text{64}^\text{3}$ and $\text{32}^\text{3}$ LUTs require 7.1 MB and 800 kB, while a GLUT-64 model occupies only 9.7 kB, resulting in compression rates of 0.14\% and 1.05\% for 75-LUT and 225-LUT settings. Such compactness makes GLUT ideal for memory-constrained applications like mobile caching or network transmission. Although ENNELUT provides the highest compression ratio, our method achieves a superior balance of fidelity and processing speed.

\begin{figure}[t!]
    \centering
    \begin{minipage}[b]{0.73\textwidth}
        \centering
        \includegraphics[width=\linewidth]{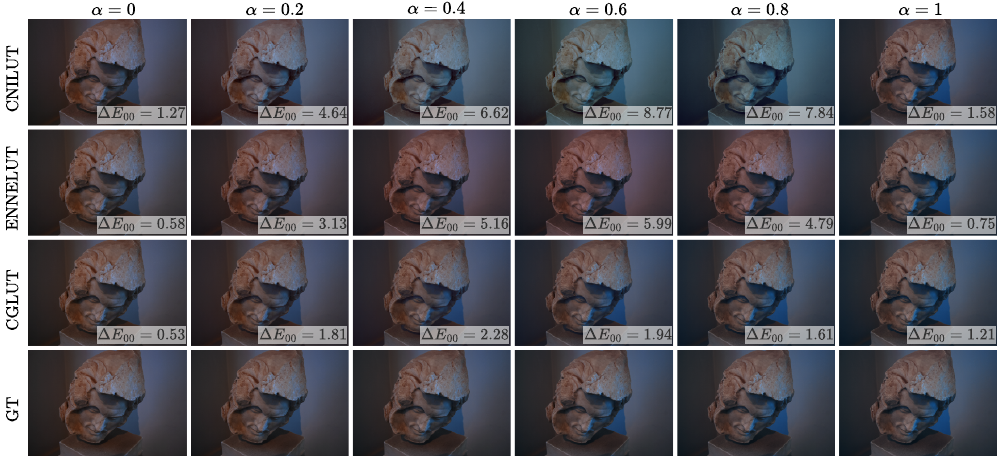}
        \caption{Qualitative comparison of style blending with different $\alpha$. Rows 1–4 illustrate the results of CNILUT (256$\times$3), ENNELUT ($L_2$), CGLUT-32L (Shared Geo.), and image-space blending (GT), respectively.}
        \label{fig:blend}
    \end{minipage}
    \hfill 
    \begin{minipage}[b]{0.24\textwidth}
        \centering
        \includegraphics[width=\linewidth]{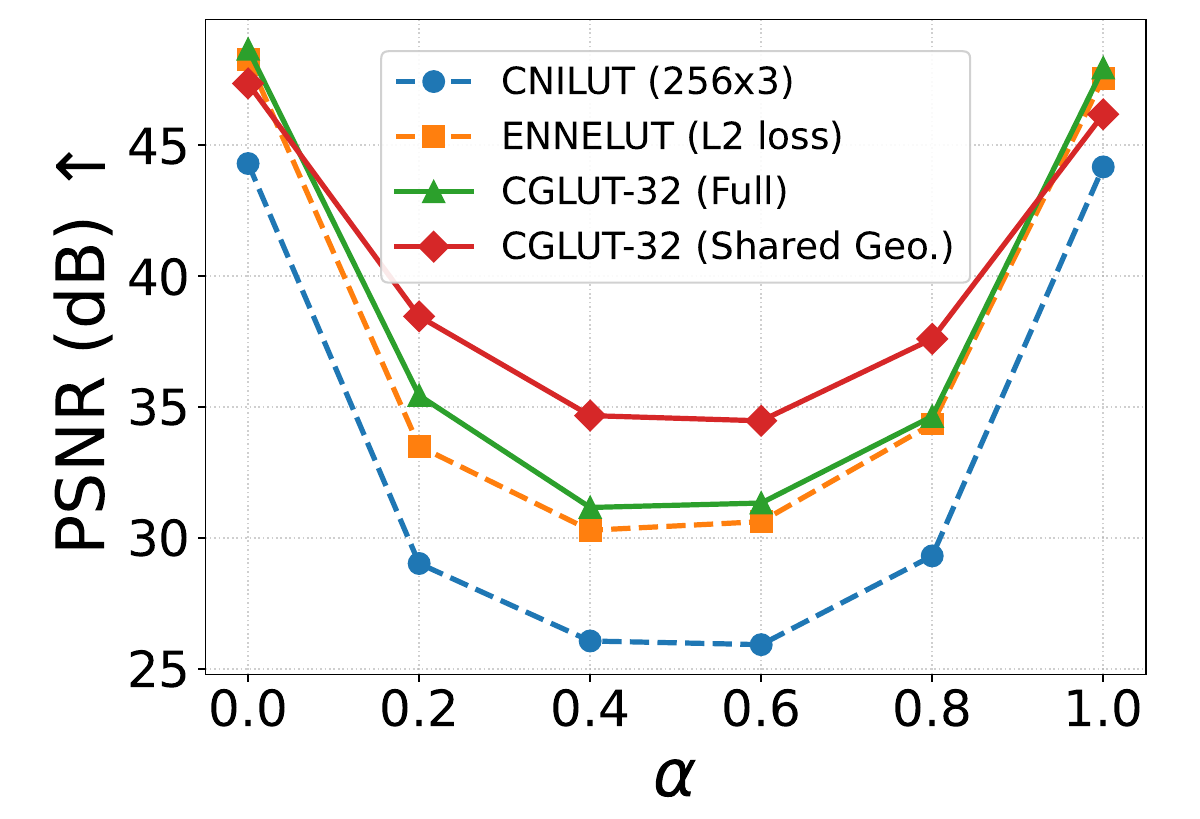}
        \includegraphics[width=\linewidth]{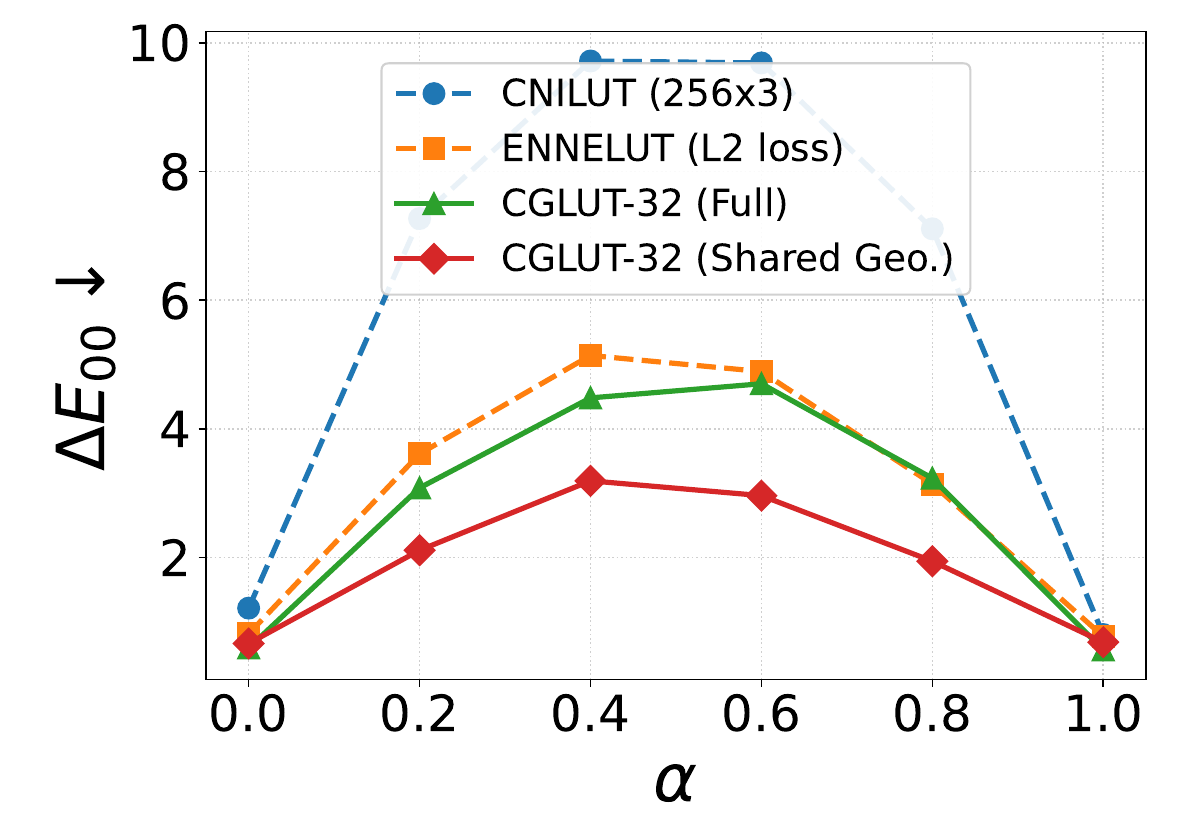}
        \caption{Quantitative comparison of blending with different $\alpha$.}
        \label{fig:blend_curve}
    \end{minipage}
\end{figure}

\noindent \textbf{Blending LUTs.}
\Cref{fig:blend} illustrates a qualitative comparison of style blending across different models. By interpolating the learned style embeddings, CGLUT achieves high-quality transitions that are consistent with direct linear blending in image space.  In contrast, although CNILUT and ENNELUT perform comparably on the two endpoint styles, their intermediate results differ: CNILUT exhibits significant color shifts, whereas ENNELUT produces smaller shifts. This observation is further validated in \Cref{fig:blend_curve}, which evaluates the blended LUTs on 100 images from the MIT5K dataset \cite{bychkovsky2011learning}. Both CGLUT variants, using fully generated parameters and shared geometry, outperform ENNELUT and CNILUT in terms of PSNR and $\Delta E$, with the shared-geometry variant achieving the best overall performance. Additional blending results are presented in \Cref{sup_sec:blend} of the Appendix.

\noindent \textbf{GLUT Editing.}
\Cref{fig:editing_main} shows six examples of editing trained GLUT representations using our proposed method. 
In the first row, adjusting $\mathbf{c}$ shifts the sky toward a darker blue or the ground toward a darker orange, while leaving the rest of the color transformation largely unchanged. In the second row, varying $K$ controls the locality of the edit within the RGB cube. When $K=1$, the correction is concentrated near $\mathbf{c}_\text{in}$, as only the most influential Gaussian primitive is updated. When $K=16$, the edit spreads more globally as a larger set of Gaussians contributes to the correction. In the third row, with $\mathbf{c}$ and $K$ fixed, increasing $s$ progressively shifts the orangish leaves toward purple, demonstrating that $s$ provides intuitive, continuous control over the edit magnitude.

We emphasize that GLUT editing is not designed for semantically or spatially aware manipulation. Rather, it provides a real-time mechanism to adjust the \emph{mood} or \emph{style} of an existing color transformation without retraining. We analyze the locality of GLUT editing in \Cref{sup_sec:locality} and describe an interactive editing tool in \Cref{sup_sec:editing} of the Appendix.

\begin{figure}[t]
    \includegraphics[width=\linewidth]{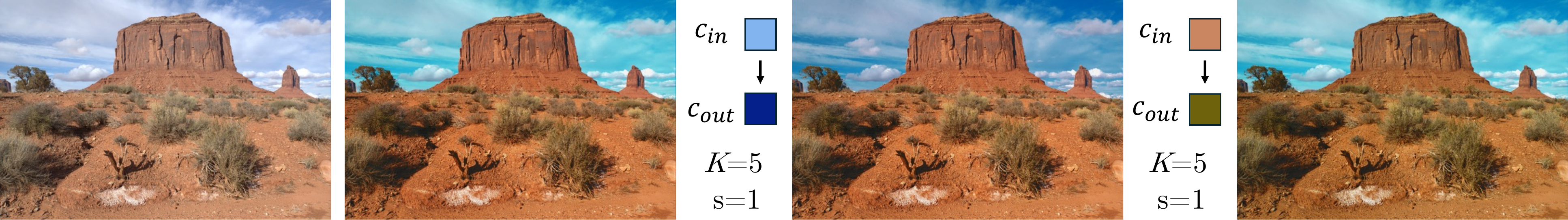}
    \includegraphics[width=\linewidth]{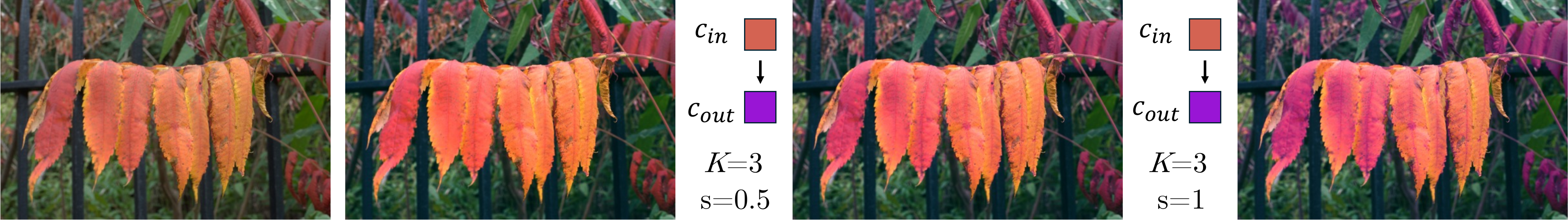}
    \includegraphics[width=\linewidth]{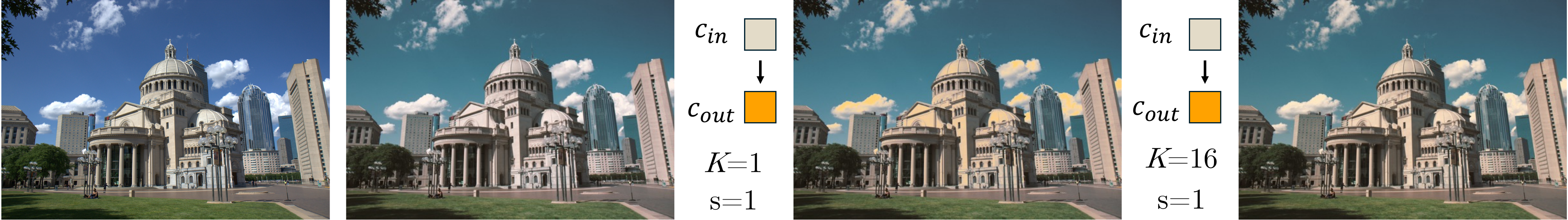}
    \makebox[0.21\textwidth]{\footnotesize Input}
    \makebox[0.21\textwidth]{\footnotesize GLUT processed}
    \makebox[0.07\textwidth]{\footnotesize }
    \makebox[0.21\textwidth]{\footnotesize Editing 1}
    \makebox[0.06\textwidth]{\footnotesize }
    \makebox[0.21\textwidth]{\footnotesize Editing 2}
    \caption{Qualitative results of editing GLUT representations. Each row shows an input image, a retouched image by GLUT, and two edited versions obtained with different color constraints $\mathbf{c} = (\mathbf{c}_\text{in}, \mathbf{c}_\text{out})$, number of Gaussians to edit $K$, and strength $s$.}
    \label{fig:editing_main}
\end{figure}

\noindent \textbf{Ablation Studies.}
We conduct ablation studies with the GLUT-32 model on the 75-LUT Hald dataset. We evaluate the number of Gaussian primitives, different local transformation formulations, key architectural components, loss terms, and training strategies, including hard sample mining and initialization. The results in \Cref{sup_sec:ablation} of the Appendix assess the contribution of each design choice and demonstrate the effectiveness of our proposed framework.

\section{Concluding Remarks}

We propose GLUT, a continuous color look-up table representation based on explicit Gaussian primitives. Unlike previous discrete or implicit methods, GLUT provides a compact, interpretable, and fully differentiable framework for global color grading and fine-grained manipulation. Our results show that GLUT achieves high color fidelity while offering remarkable computational efficiency. We further introduce CGLUT, a conditional variant that represents multiple LUTs within a unified model and supports smooth blending between color styles. By replacing rigid grids with optimizable, spatially localized primitives, GLUT offers a flexible approach to color manipulation. We envision this representation serving as a building block for a broader range of applications, including real-time ISP tuning and interactive artistic grading.

\clearpage

\begin{ack}
This work was supported by Grant PID2024-162555OB-I00 funded by MCIN/AEI/10.13039/501100011033 and by ERDF, EU, AIA2025-163919-C52 funded by MCIN/AEI/10.13039/501100011033, the grant Catedra ENIA UAB-Cruïlla (TSI-100929-2023-2) from the Ministry of Economic Affairs and Digital Transformation of Spain, and by the Generalitat de Catalunya CERCA Program. Danna Xue and Javier Vazquez-Corral also acknowledge the 2025 Leonardo Grant for Scientific Research and Cultural Creation from the BBVA Foundation. The BBVA Foundation accepts no responsibility for the opinions, statements and contents included in the project and/or the results thereof, which are entirely the responsibility of the authors. David Serrano-Lozano also acknowledges the FPI grant from the Spanish Ministry of Science and Innovation (PRE2022-101525). Shaolin Su was supported by the HORIZON MSCA Postdoctoral Fellowship funded by the European Union (project number 101152858). The authors declare no competing interests.
\end{ack}

{\small
\bibliographystyle{plain}
\bibliography{main}
}

\newpage

\appendix
\section*{Appendix}

\input{supplementary}


\newpage

\end{document}

%% file: supplementary.tex
In this appendix, we provide the statement about limitations and broader impacts, \etc, in \Cref{sup_sec:limit}, additional implementation details in \Cref{sup_sec:imple}, followed by additional experimental results and ablation studies in \Cref{sup_sec:exp}. Furthermore, we offer a step-by-step user instruction for our interactive GLUT-based LUT editing tool in \Cref{sup_sec:editing}.

\section{Statements}
\label{sup_sec:limit}

\noindent \textbf{Limitations and Discussion.}
While GLUT demonstrates strong performance in color transformation tasks, several limitations should be acknowledged. 
First, GLUT is designed to improve the efficiency and interpretability of neural color LUTs, which are inherently limited to global color transformations, where the output depends solely on the input RGB values. Although GLUT currently focuses on global color transformations, it could potentially be extended to context-dependent tasks, such as spatially varying image retouching, with further research, as would be required for conventional LUTs.
Second, for both GLUT and CGLUT, the number of Gaussian primitives $N$ is set as a fixed hyperparameter prior to training, yet the optimal $N$ may vary across different color styles and datasets. While we provide empirical guidelines, automatically determining the optimal $N$ during training remains an open problem. Techniques analogous to adaptive density control in 3D Gaussian Splatting could potentially be adapted for this purpose. 
Third, although GLUT supports localized LUT editing, its locality depends on the spatial extent of the neighboring Gaussian primitives. Extending the editing framework to support finer-grained control, for example by adding Gaussian primitives in a specific region or fine-tuning additional Gaussian parameters, is a potential direction for future work.

\noindent \textbf{Broader Impacts.}
This work advances the efficiency and interpretability of color transformation technology, with potential positive impacts across several domains. In professional photography and cinematography, GLUT offers a compact and editable representation of color grading styles, enabling efficient storage, transmission, and application of color presets in large-scale media production pipelines. The interactive GLUT Editor facilitates intuitive color correction workflows, potentially making professional-grade color grading tools more accessible for independent creators. In computational photography, the learned continuous LUT representation could improve camera ISP calibration and cross-device color matching, benefiting consumer imaging applications. The compact model size, which can be substantially smaller than that of traditional LUTs, makes color transformation more efficient on resource-constrained devices such as smartphones and embedded systems, broadening access to professional color science tools. 

We do not foresee significant negative societal impacts specific to this work. Color transformation is a well-established component of standard media production workflows, and GLUT improves its efficiency rather than introducing qualitatively new capabilities. However, as with any image processing technology, efficient color transformations could potentially be used in image manipulation or deepfake production pipelines. We emphasize that GLUT is designed for color transformation rather than content generation, and its application to deceptive media manipulation would require additional components beyond the scope of this work.

\noindent \textbf{Ethical Statement.}
This work addresses the technical problem of efficient color transformation representation and does not involve human subjects, personally identifiable information, or sensitive data collection. All experiments were conducted on publicly available image datasets. The proposed GLUT Editor is designed to assist professional colorists and photographers in their creative workflows, with all designs prioritizing transparency and user control over the color transformation process. The explicit Gaussian representation provides interpretable parameters, including the location and spatial extent of each primitive in RGB space, allowing users to inspect and directly edit the color transformation. We have released all code, pretrained models, and training configurations to support open and reproducible research in color science and image processing.

\noindent \textbf{Reproducibility Statement.}
To facilitate reproducibility, we have released the complete source code, including model implementations, training and evaluation scripts, pretrained model checkpoints, and the GLUT Editor interface at \href{https://github.com/CVC-Color/glut}{https://github.com/CVC-Color/glut}. Hyperparameter configurations for all reported experiments are provided in the appendices. All experiments reported in this paper were conducted with fixed random seeds. The datasets used in this work are either publicly available or accompanied by detailed instructions for data preparation. The GLUT Editor is implemented as a self-contained browser application (HTML/JavaScript, CPU-only) that runs without installation.

\noindent \textbf{Future Work.}
A possible future direction involves integrating spatial or semantic features into the Gaussian parameter generator, extending CGLUT into a spatially varying representation that enables content-aware color manipulation while preserving the efficiency of explicit primitives. Moreover, the compact style embeddings of CGLUT provide a natural interface for multimodal control. Another possible direction involves leveraging vision-language models to integrate text prompts into GLUT parameter generation. This will enable intuitive, language-driven color grading, thereby further expanding the applications of image color editing.

\section{Implementation Details}
\label{sup_sec:imple}

\subsection{Training and Evaluation}
\label{sup_sec: train}
\noindent \textbf{Initialization.} For GLUT initialization, Gaussian means are initialized by distributing them uniformly on a regular grid covering the RGB cube $[0,1]^3$. Covariances are initialized isotropically with a scale of $\sigma=0.15$ through logarithmic Cholesky parameters. Opacities are initially set to 1.0, and the affine color transforms are initialized as identity matrices with zero bias.

\noindent \textbf{Training of CGLUT.} Since CGLUT comprises multiple components with different roles, we apply a lower learning rate (0.1$\times$ the base rate) to the style embeddings and shared geometry parameters, while the generator (shared feature encoder and parameter heads) uses the base learning rate of $10^{-3}$. We set $\epsilon=10^{-6}$ in \Cref{eq:weights} and \Cref{eq:sparsity} for numerical stability.

\noindent \textbf{Hard Sample Mining.} We employ a curriculum hard-mining strategy to guide the model toward learning complex and high-error color transformations. Specifically, from epoch 5 to 20, the proportion of samples with the highest $L_1$ errors selected for mining increases linearly from 10\% to 40\%. This strategy encourages the model to focus progressively on challenging mappings and reduces its tendency to prioritize easy-to-fit samples.

\noindent \textbf{Datasets.} As is common practice, Hald images~\cite{haldwebsite} are widely used for representing LUTs and as training samples for neural LUT methods~\cite{conde2024nilut, zehtab2025efficient}. We uniformly sample the full 8-bit RGB space to construct a $128^3$ training set, reserving the remaining colors for evaluation to verify that the model learns a continuous representation. The resulting mappings are reshaped into Hald images of resolutions 1024$\times$2048$\times$3 and 3584$\times$4096$\times$3, respectively. By applying different LUTs to the original Hald image, we obtain output Hald images of the same resolution but with different color styles. These Hald image pairs are used for training and testing. For evaluation on Hald images, we report PSNR, $\Delta E_{00}$, and $\Delta E_{76}$.

To assess color transformation quality in more general scenarios, we also test on 100 natural images from the MIT-Adobe FiveK (MIT5K) dataset~\cite{bychkovsky2011learning} (images \#4501 to \#4600). We use the images processed by traditional LUTs (\texttt{.cube} files) as ground truth, and compare them with images processed by applying the learned NILUT, CNILUT, ENNELUT, GLUT and CGLUT models. These natural images are not used for training. The color fidelity is evaluated using PSNR, SSIM, LPIPS, $\Delta E_{00}$, and $\Delta E_{76}$.

\subsection{CGLUT Architecture}
\label{sup_sec:archi}
\begin{figure}[ht]
    \centering
    \includegraphics[width=0.9\linewidth]{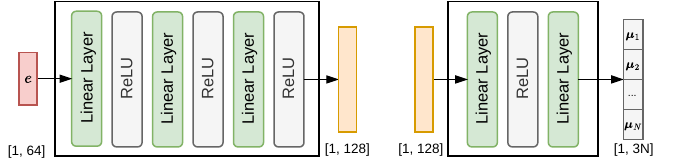}
    \makebox[0.52\linewidth]{\footnotesize (a) Shared encoder}
    \makebox[0.46\linewidth]{\footnotesize (b) Mean head}
    \caption{Architecture of the Gaussian Parameter Generator. A style embedding is first projected through a shared MLP encoder to extract common latent features. Subsequently, the representation is branched into task-specific heads to predict different parameters of the Gaussian primitives.}
    \label{fig:architecture}
\end{figure}

\Cref{fig:architecture} illustrates the architecture of the Gaussian parameter generator in our CGLUT. Given a 64-dimensional style embedding, a shared encoder, consisting of three linear layers with 128 (64 for ``small'' setup) hidden units each, extracts a latent feature representation. This feature is then fed into multiple task-specific heads to predict the Gaussian parameters. As shown in \Cref{fig:architecture} (b), the head responsible for mean values $\boldsymbol{\mu}$ comprises two linear layers: an initial layer with 128 (64 for ``small'' setup) units, followed by an output layer whose dimension is determined by the number of Gaussians $N$. For instance, to predict 3D mean values, the output dimension is $3N$. Other parameter heads follow a similar structure with adjusted output dimensions tailored to their respective parameters. For the specific case of the global affine transform, since each GLUT incorporates a single global affine transformation, the corresponding head outputs 12 parameters (9 for the matrix and 3 for the bias). All the parameter heads have the same structure as the mean head with two linear layers, except for the local color head, which has three linear layers to learn more complicated local color transformation parameters. All intermediate layers are interleaved with ReLU activations to ensure sufficient representational capacity.

In the \textit{Shared Geometry} configuration of CGLUT, the mean and covariance values are shared across all styles. Consequently, the mean and covariance heads are replaced by a set of globally learnable parameters, similar to the single-GLUT setting, while only the opacities and color transformations are dynamically generated.

\textbf{Scalability of CGLUT.}
The output dimensionality of the CGLUT generator does not grow with the number of styles. In each forward pass, the generator maps a single style embedding $\mathbf{e}_l$ to the parameters of one GLUT instance, which has $22N+12$ values, or $13N+12$ values per style under shared geometry. This dimensionality is the same whether CGLUT is trained on 75 or 225 styles. Adding styles increases the number of training samples and style embeddings, but each prediction remains a fixed-size regression. Consequently, the model size also grows slowly. Our largest configuration, CGLUT-64-Large trained on 225 styles, has 338K parameters in total, of which 323.6K belong to the shared generator and the remaining 14.4K to the 64-dimensional style embeddings. Each additional style therefore adds only one 64-dimensional embedding, increasing the total parameter count by just 64.

\subsection{CUDA Acceleration of GLUT}
To accelerate inference, we implement a custom CUDA kernel for the GLUT forward pass. The kernel fuses the Gaussian weight computation, local color transformation, and weighted color blending into a single GPU kernel, eliminating intermediate memory reads and writes that would otherwise dominate the runtime. The thread organization assigns one pixel per thread along the first dimension and distributes the $N$ Gaussian primitives across threads along the second dimension within each thread block. The block dimensions are set to $(32, 32)$, aligning the first dimension with the warp size of modern NVIDIA GPUs.

Gaussian parameters shared across all pixels are precomputed once per forward pass and cached in registers and shared memory, avoiding redundant recomputation for each pixel. For training, the standard PyTorch autograd path is retained to preserve gradient flow. The custom CUDA kernel is activated exclusively during inference.

\subsection{Classical Baselines}
\label{sup_sec:classic}

All baselines in \Cref{tab:classic} are fitted on the training colors of the 75-LUT and 225-LUT Hald datasets and evaluated on the held-out test colors, following the protocol in \Cref{sec:exp_setup}. Except for root-polynomial regression~\cite{finlayson2015color}, all baselines are configured to match the 716-parameter budget of GLUT-32. The baselines are divided into two groups. The learned baselines, namely the RBF network~\cite{moody1989fast} and GLLiM-GD~\cite{deleforge2015high}, are trained end-to-end with GLUT's pipeline. 
They share the objective $\mathcal{L}_\text{rec} + \lambda_{hc}\mathcal{L}_{hc}$, without the sparsity regularizer, as these models do not involve opacity parameters. The remaining baselines are fitted in closed form or using their official implementations. The grid-based~\cite{fumoto1997extraction,kasson1995performing} and keypoint-based baselines~\cite{tschumperle2020reconstruction} require a reference LUT defined on a complete regular grid, which we obtain from the training Hald image covering the full $128^3$ RGB grid.

\paragraph{Root-polynomial regression~\cite{finlayson2015color}.}
This model maps each color through a fixed global feature expansion followed by a linear transformation. We use the second-order expansion $(r, g, b, \sqrt{rg}, \sqrt{gb}, \sqrt{rb})$ without an intercept, following the original formulation, together with a 3$\times$6 matrix, giving 18 parameters. The matrix is fitted by least squares over all training colors. We do not match the budget for this baseline. It is a single global mapping originally designed for device characterization, so it has no mechanism to allocate capacity locally.

\paragraph{RBF network~\cite{moody1989fast}.}
We use a normalized RBF network with isotropic Gaussian kernels and constant outputs:
\begin{equation}
f(\mathbf{x}) = \sum_i \frac{\varphi_i(\mathbf{x})}{\sum_j \varphi_j(\mathbf{x})}\,\mathbf{c}_i + \mathbf{c}_0,
\qquad
\varphi_i(\mathbf{x}) = \exp\!\Big(-\frac{\|\mathbf{x}-\boldsymbol{\mu}_i\|^2}{2\sigma_i^2}\Big).
\end{equation}
Each unit stores a center $\boldsymbol{\mu}_i \in \mathbb{R}^3$, a bandwidth $\sigma_i \in [0,1]$, and an output color $\mathbf{c}_i\in \mathbb{R}^3$, which amounts to 7 parameters, and a shared bias $\mathbf{c}_0 \in \mathbb{R}^3$ adds 3 parameters. We use 102 units, yielding 717 parameters to approximately match the budget of GLUT-32. Compared with GLUT, this model uses constant rather than affine local transformations, isotropic rather than anisotropic Gaussian kernels, and a shared bias rather than a global affine transformation. It has no opacity parameters. The centers are initialized on the same regular grid as GLUT, and each output color is initialized to its corresponding center, making the initial mapping close to the identity.

\paragraph{GLLiM (GD/EM)~\cite{deleforge2015high}.}
GLLiM models the mapping as a Gaussian-gated mixture of local affine experts:
\begin{equation}
f(\mathbf{x}) = \sum_{k} r_k(\mathbf{x})\,(\mathbf{A}_k\mathbf{x}+\mathbf{b}_k),
\qquad
r_k(\mathbf{x}) = \frac{\pi_k\,\mathcal{N}(\mathbf{x};\mathbf{c}_k,\boldsymbol{\Gamma}_k)}{\sum_j \pi_j\,\mathcal{N}(\mathbf{x};\mathbf{c}_j,\boldsymbol{\Gamma}_j)}.
\end{equation}
Each component has 22 parameters: a mean, a full covariance, a mixing weight, and an affine expert, which equals the per-primitive count of GLUT. GLLiM differs from GLUT in three aspects: its mixing weights lie on the probability simplex, it has no global affine term, and it does not use the sparsity regularizer. We use 33 components to match the budget, giving 726 parameters. Both variants use the same initialization as GLUT, with means on a regular grid, isotropic covariances with scale $\sigma=0.15$, uniform mixing weights, and identity affine experts. \textbf{GLLiM (GD)} trains this predictor end-to-end with our training pipeline. It parameterizes each covariance by its Cholesky factor and the mixing weights by a softmax over learnable logits. \textbf{GLLiM (EM)} fits the generative model of~\cite{deleforge2015high} using expectation-maximization with closed-form updates. We run 500 EM iterations on all training color pairs.

\paragraph{Keypoint diffusion~\cite{tschumperle2020reconstruction}.}
We use the official G'MIC implementation of Tschumperlé \etal~\cite{tschumperle2020reconstruction}. Compression iteratively inserts keypoints where the reconstruction error is largest, and decompression reconstructs a dense LUT from the keypoints by multi-scale anisotropic diffusion. Each keypoint stores an input position and an output color, which amounts to 6 parameters. We cap the number of keypoints at 120, giving 720 parameters. We reconstruct the LUT at the training resolution of $128^3$ in sRGB space and query it with trilinear interpolation.

\paragraph{Sparse grid LUTs~\cite{fumoto1997extraction,kasson1995performing}.} The largest regular grid within the budget of GLUT-32 has $6^3$ nodes, giving $6^3\times3=648$ parameters. We obtain the node values by downsampling the reference LUT to this grid and query the grid with two interpolation schemes. Tetrahedral interpolation~\cite{kasson1995performing} splits each cell into six tetrahedra that share its main diagonal and interpolates linearly within the tetrahedron containing the query, which yields a continuous, piecewise-linear mapping. Tricubic interpolation~\cite{fumoto1997extraction} fits a tensor-product cubic B-spline that passes through the node values, which yields a twice continuously differentiable mapping. We do not further optimize the node values in this comparison. Fitting them to the training colors, as in lattice regression~\cite{garcia2009lattice}, improves the PSNR on the 75-LUT dataset to 37.79 dB for tetrahedral and 37.40 dB for tricubic interpolation, which remains below that of GLUT-32.

\section{Additional Experiment Results}
\label{sup_sec:exp}

\subsection{Single-LUT Results}
\label{sup_sec:glut}

\subsubsection{Single LUT Fitting}
\Cref{sup_tab:glut-hald} compares the performance of our proposed GLUT against NILUT on 225-LUT Hald images and 100 natural images from the MIT5K dataset \cite{bychkovsky2011learning}. GLUT achieves superior PSNR and $\Delta E$ while maintaining a significantly smaller parameter footprint and lower computational cost. Notably, the color fidelity on this set is higher than that observed on the 75-LUT dataset (\Cref{tab:glut-hald} in the main paper). This is because the LUTs in the 75-LUT set feature a higher resolution ($\text{64}^\text{3}$) than those in the 225-LUT set ($\text{32}^\text{3}$ or $\text{33}^\text{3}$ ). Higher-resolution LUTs can represent finer color variations, potentially making accurate reconstruction more challenging.

\begin{table}[ht]
\centering
\caption{Quantitative comparison of single-LUT modeling on Hald images and natural images (225 LUTs).}
\label{sup_tab:glut-hald}
\resizebox{\columnwidth}{!}{%
\begin{tabular}{l|ccc|ccccc|c|c}
\toprule
\multicolumn{1}{c|}{\multirow{2}{*}{Method}} & \multicolumn{3}{c|}{Hald images} & \multicolumn{5}{c|}{Natural images} & \multirow{2}{*}{\#Params} & \multirow{2}{*}{GFLOPs} \\
\multicolumn{1}{c|}{} & PSNR$\uparrow$ & $\Delta E_{00}\downarrow$ & $\Delta E_{76}\downarrow$ & PSNR$\uparrow$ & SSIM$\uparrow$ & LPIPS$\downarrow$ & $\Delta E_{00}\downarrow$ & $\Delta E_{76}\downarrow$ &  &  \\ \midrule
NILUT (64$\times$2)~\cite{conde2024nilut} & 42.49 & 0.76 & 1.45 & 41.98 & 0.987 & 0.0109 & 2.14 & 2.81 & 8,771 & 4.60 \\
NILUT (128$\times$2)~\cite{conde2024nilut} & 44.77 & 0.57 & 1.08 & 44.21 & 0.992 & 0.0066 & 1.61 & 2.04 & 33,923 & 17.78 \\
GLUT-16 & 46.69 & 0.45 & 0.88 & 46.23 & 0.994 & 0.0037 & 1.06 & 1.44 & 364 & 0.25 \\
GLUT-32 & \textit{50.42} & \textit{0.30} & \textit{0.58} & \textit{48.74} & \textit{0.997} & \textit{0.0020} & \textit{0.75} & \textit{0.99} & 716 & 0.49 \\
GLUT-64 & \textbf{52.73} & \textbf{0.22} & \textbf{0.43} & \textbf{50.56} & \textbf{0.998} & \textbf{0.0011} & \textbf{0.58} & \textbf{0.76} & 1,420 & 0.98 \\\bottomrule
\end{tabular}%
}
\end{table}

\begin{figure}[h!]
   \centering
    \includegraphics[width=1\linewidth]{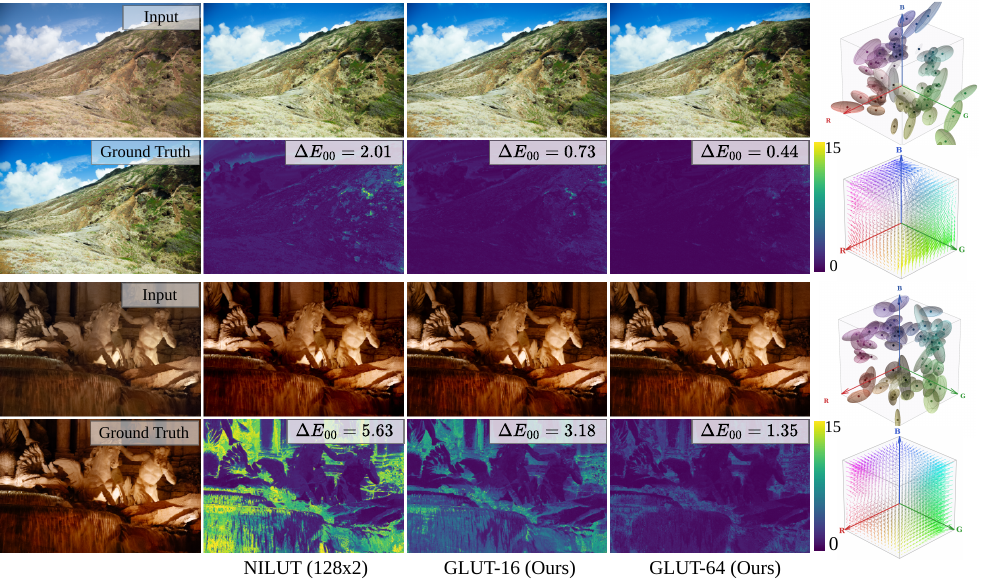}
    \caption{Qualitative results on images from the MIT5K~\cite{bychkovsky2011learning}. The first column presents the input and the retouched image with a LUT. The top row shows the retouched version of the input image of NILUT (128$\times$2) \cite{conde2024nilut}, GLUT-16, and GLUT-64. Under each image, we present the $\Delta E_{00}$ error map. The last column shows the visualization of our GLUT-64 representation and its corresponding LUT. Each Gaussian primitive is visualized using its learned mean values.}
    \label{fig:glut-image1}
\end{figure}

\begin{figure}[ht]
    \centering
    \includegraphics[width=0.95\linewidth]{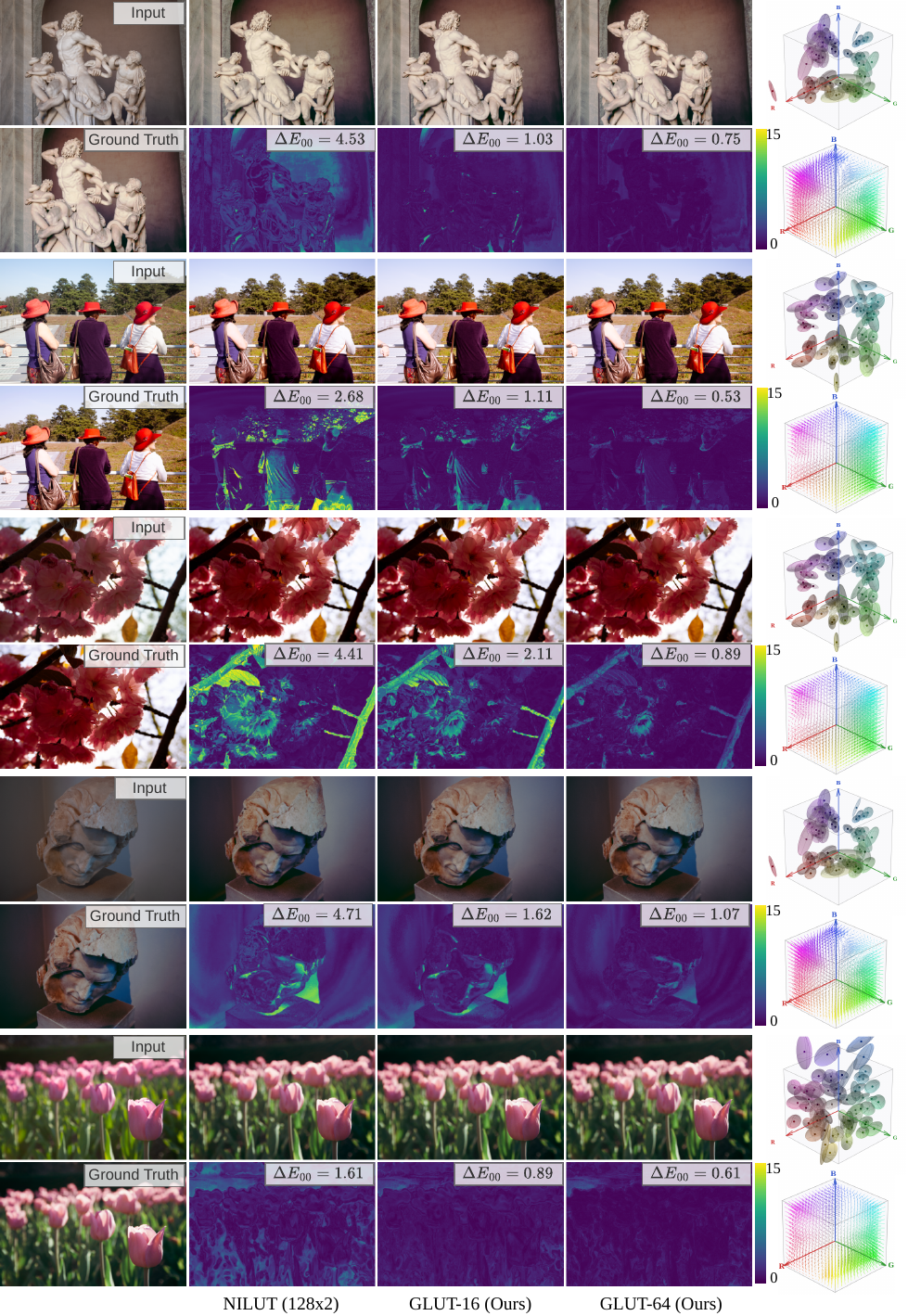}
    \caption{Qualitative results on images from the MIT5K~\cite{bychkovsky2011learning}. The first column presents the input and the retouched image with a LUT. The top row shows the retouched version of the input image of NILUT(128$\times$2) \cite{conde2024nilut}, GLUT-16, and GLUT-64. Under each image, we present the $\Delta E_{00}$ error map. The last column shows the visualization of our GLUT-64 representation and its corresponding LUT. Each Gaussian primitive is visualized using its learned mean value. }
    \label{fig:glut-image}
\end{figure}

\Cref{fig:glut-image1} and \Cref{fig:glut-image} show the qualitative comparison between images retouched with the same color styles with NILUT and GLUT. As the number of Gaussian primitives $N$ increases, the model's ability to capture complex color mappings improves consistently, leading to lower reconstruction errors. For both NILUT and GLUT, the residual errors are mainly concentrated in darker intensity ranges, such as shadowed regions. This suggests that representing high-contrast transitions in low-luminance areas remains a common challenge for both implicit and explicit neural color representations. The last columns of \Cref{fig:glut-image1} and \Cref{fig:glut-image} also visualize the GLUT-64 representation alongside the ground-truth LUT. Notably, the Gaussian distribution exhibits a structural correspondence with the target LUT, with more Gaussians located in regions associated with larger color shifts.

\subsubsection{Comparison with INRetouch}
\label{sup_sec:inretouch}
Since INRetouch \cite{elezabi2024inretouch} primarily targets preset-based color style transfer, its retouched outputs involve spatially varying transformations, where identical input RGB values can be mapped to different output values depending on their spatial coordinates. In contrast, our GLUT focuses on globally consistent color transformations. Similar to our single-LUT model GLUT, each INRetouch model is trained specifically for a single preset. Given that INRetouch's architecture is designed to capture contextual information from natural images, training it on Hald images, which are merely tabular representations of the RGB color space, would be unsuitable. To ensure an appropriate comparison, we train both models using a single image pair processed by a specific LUT and subsequently evaluate them on 100 test images from MIT5K \cite{bychkovsky2011learning}. 

As shown in \Cref{tab:inretouch}, our GLUT-32 consistently outperforms INRetouch across all metrics, including PSNR, SSIM, and $\Delta E_{76}$, on seven different LUTs. Both models achieve lower performance when trained on a single image pair than when trained using a Hald image. This performance gap stems from the fact that natural images typically do not span the entire RGB color space. Consequently, the model is only supervised on a sparse subset of colors present in the scene, leading to less robust generalization in unobserved color regions compared to the exhaustive supervision provided by a Hald image.

\begin{table}[h]
\centering
\caption{Quantitative comparison of single-LUT modeling training with a single natural image pair.}
\label{tab:inretouch}
\resizebox{0.7\textwidth}{!}{
\setlength{\tabcolsep}{4pt}
\begin{tabular}{c|l|ccccccc|c}
\toprule
Metric & \multicolumn{1}{c|}{Method} & LUT1 & LUT2 & LUT3 & LUT4 & LUT5 & LUT6 & LUT7 & Average \\ \midrule
\multirow{2}{*}{PSNR$\uparrow$} & INRetouch & 34.91 & 33.32 & 34.15 & 40.16 & 33.00 & 35.81 & 30.15 & 34.50 \\
 & GLUT-32 & \textbf{39.57} & \textbf{36.55} & \textbf{38.63} & \textbf{41.18} & \textbf{39.47} & \textbf{44.53} & \textbf{35.78} & \textbf{39.39} \\ \midrule
\multirow{2}{*}{SSIM$\uparrow$} & INRetouch & 0.967 & 0.967 & 0.960 & 0.977 & 0.954 & 0.977 & 0.916 & 0.960 \\
 & GLUT-32 & \textbf{0.994} & \textbf{0.991} & \textbf{0.991} & \textbf{0.989} & \textbf{0.992} & \textbf{0.997} & \textbf{0.980} & \textbf{0.991} \\ \midrule
\multirow{2}{*}{$\Delta E_{76}\downarrow$} & INRetouch & 4.120 & 5.096 & 5.464 & 2.643 & 5.948 & 3.625 & 7.016 & 4.845 \\
 & GLUT-32 & \textbf{2.190} & \textbf{2.360} & \textbf{2.640} & \textbf{1.890} & \textbf{3.570} & \textbf{1.430} & \textbf{2.530} & \textbf{2.370} \\ 
\bottomrule
\end{tabular}%
}
\end{table}

\subsubsection{Runtime and Implementation Efficiency}

We compare the runtime of GLUT with traditional grid-based LUTs and neural LUT representations under different implementations.

\textbf{Traditional LUTs.}
\Cref{sup_tab:runtime} reports FPS of a grid-based $\text{33}^\text{3}$ 3D LUT using trilinear interpolation under three implementations. GPU texture units achieve the highest throughput across all image resolutions, making fixed-function texture sampling highly efficient for a precomputed static LUT. However, texture units are specialized hardware in the GPU graphics pipeline and are generally unavailable on neural processing units (NPUs), which primarily target tensor and matrix operations.

\textbf{Neural LUTs.}
\Cref{sup_tab:runtime} further compares NILUT and GLUT using both standard PyTorch and optimized implementations. For NILUT, we use the fully fused MLP implementation from tiny-cuda-nn~\cite{muller2022instant}, which is specifically optimized for small coordinate-based MLPs such as NILUT (128$\times$2). For GLUT, we implement a custom fused CUDA kernel that evaluates all Gaussian primitives in a single kernel launch. With standard PyTorch implementations, GLUT is slower than NILUT despite its compact representation. This is largely due to implementation overhead: NILUT primarily relies on large matrix multiplications, which are efficiently handled by optimized CUDA libraries, whereas GLUT performs multiple small operations for each Gaussian primitive, resulting in frequent kernel launches and intermediate memory accesses. Fusing these operations into a single CUDA kernel yields an approximately 80$\times$ speedup over the PyTorch implementation. With optimized implementations for both methods, GLUT achieves approximately 8$\times$ higher throughput than NILUT across all tested resolutions. These results indicate that the runtime of the standard GLUT implementation is dominated by execution overhead from unfused operations, and that GLUT's structured computation is well suited to kernel fusion.

\begin{table}[h]
\centering
\caption{Runtime comparison of traditional and neural LUT representations under different implementations. FPS is reported at different image resolutions.}
\label{sup_tab:runtime}
\resizebox{0.6\columnwidth}{!}{%
\begin{tabular}{c|c|ccc}
\toprule
Method & Backend & $\text{512}^\text{2}$ & HD & 4K \\ \midrule
3D LUT & CPU Numba & 3489 & 759 & 148 \\
3D LUT & CUDA PyTorch & 5936 & 1025 & 265 \\
3D LUT & GPU texture unit & 26673 & 5577 & 1510 \\ \midrule
NILUT(128×2)~\cite{conde2024nilut} & CUDA PyTorch & 295 & 39 & 8 \\
NILUT(128×2)~\cite{conde2024nilut} & tiny-cuda-nn & 822 & 111 & 28 \\ \midrule
GLUT-32 & CUDA PyTorch & 79 & 10 & 3 \\
GLUT-32 & Fused CUDA kernel & 6170 & 891 & 226 \\ \bottomrule
\end{tabular}%
}
\end{table}

In \Cref{tab:runtime} in the main paper, we only report CNILUT and ENNELUT with their PyTorch implementations, since tiny-cuda-nn gives CNILUT no measurable speedup in our tests. ENNELUT injects its condition inside each block, uses the LipSwish activation, and has hidden layers of varying width, none of which the fully fused MLP of tiny-cuda-nn supports.


\subsection{Multiple-LUT Results}

\subsubsection{CGLUT Results on Natural Images}
\label{sup_sec:glut_image}

\Cref{sup_tab:cglut_image} presents the performance evaluation on 100 test images from the MIT5K dataset \cite{bychkovsky2011learning}. Our CGLUT consistently outperforms CNILUT \cite{conde2024nilut} and ENNELUT \cite{zehtab2025efficient} across both 7-LUT, 75-LUT, and 225-LUT configurations. The testing results on natural images align with the results on Hald images shown in \Cref{tab:cglut} in the main paper.

\begin{table}[ht]
\centering
\caption{Quantitative comparison of multiple-LUT modeling on the natural images from MIT5K~\cite{bychkovsky2011learning}.}
\label{sup_tab:cglut_image}
\setlength{\tabcolsep}{3pt}
\resizebox{\textwidth}{!}{%
\begin{tabular}{l|ccccc|ccccc|ccccc}
\toprule
\multicolumn{1}{c|}{\multirow{2}{*}{Method}} & \multicolumn{5}{c|}{7 LUTs} & \multicolumn{5}{c|}{75 LUTs} & \multicolumn{5}{c}{225 LUTs} \\
\multicolumn{1}{c|}{} & PSNR$\uparrow$ & SSIM$\uparrow$ & LPIPS$\downarrow$ & $\Delta E_{00}\downarrow$ & $\Delta E_{76}\downarrow$ & PSNR$\uparrow$ & SSIM$\uparrow$ & LPIPS$\downarrow$ & $\Delta E_{00}\downarrow$ & $\Delta E_{76}\downarrow$ & PSNR$\uparrow$ & SSIM$\uparrow$ & LPIPS$\downarrow$ & $\Delta E_{00}\downarrow$ & $\Delta E_{76}\downarrow$ \\ \midrule
CNILUT (128$\times$3)~\cite{conde2024nilut} & 40.12 & 0.995 & 0.019 & 1.74 & 2.21 & 39.78 & 0.987 & 0.027 & 2.45 & 2.95 & 38.93	&0.984	&0.028	&2.77	&3.58  \\
CNILUT (256$\times$3)~\cite{conde2024nilut} & 44.24 & \textit{0.998} & 0.011 & 1.18 & 1.48 & 42.75 & 0.992 & 0.020 & 1.84 & 2.16 & 41.98 & 0.990 &0.019	& 2.07	&2.65  \\ \midrule
ENNELUT ($\Delta E$)~\cite{zehtab2025efficient} & 45.83 & \textit{0.998 }& 0.008 & 0.79 & 1.03 & 43.23 & 0.993 & 0.016 & 1.26 & 1.52 & 40.50 & 0.989 & 0.020 & 2.02 & 2.61 \\
ENNELUT ($L_2$)~\cite{zehtab2025efficient} & 47.93 & \textit{0.998} & 0.008 & 0.79 & 0.96 & 43.43 & \textit{0.994} & 0.018 & 1.52 & 1.76 & 40.47 & 0.989 & 0.024 & 2.41 & 3.04 \\ \midrule
CGLUT-32 (Small) & 46.42 & \textit{0.998} & 0.007 & 0.70 & 0.92 & 46.61 & \textbf{0.998} & \textit{0.011} & 0.88 & 1.07 & 46.52 & 0.995 & 0.009 & 1.03 & 1.39 \\
CGLUT-32 (Large) & 48.31 & \textit{0.998} & 0.005 & 0.58 & 0.76 & 47.18 & \textbf{0.998} & \textit{0.011} & 0.82 & 0.99 & \textit{48.74} & \textit{0.997} & \textit{0.006} & \textit{0.76} & \textit{1.01} \\
CGLUT-64 (Small) & \textit{48.84} & \textit{0.998} & \textit{0.004} & \textit{0.55} & \textit{0.72} & \textit{48.38} & \textbf{0.998} & \textbf{0.010} & \textit{0.73} & \textit{0.87} & 48.53 & \textit{0.997} & \textit{0.006} & 0.78 & 1.03 \\
CGLUT-64 (Large) & \textbf{49.84} & \textbf{0.999} & \textbf{0.003} & \textbf{0.45} & \textbf{0.59} & \textbf{48.69} & \textbf{0.998} & \textbf{0.010} & \textbf{0.70} & \textbf{0.84} & \textbf{50.61} & \textbf{0.998} & \textbf{0.004} & \textbf{0.57} & \textbf{0.75} \\ \bottomrule
\end{tabular}%
}
\end{table}

\subsubsection{Blending LUTs}
\label{sup_sec:blend}

\Cref{tab:blending_7luts} presents the numerical blending results for all possible pairwise LUT combinations under the 7-LUT. Specifically, we evaluate $C^{2}_{7}=21$ unique style pairs, with metrics averaged over 100 test images from the MIT-Adobe FiveK dataset \cite{bychkovsky2011learning}.

\Cref{tab:blending_7luts} illustrates a common trend across all methods, where blending qualities are better when alpha is near 0 or 1. 
Among the models, CGLUT with fully generated parameters (Full) achieves better performance than CNILUT and ENNELUT. Notably, using CGLUT with shared geometry (Shared Geo.) further enhances the results.
This demonstrates that our Gaussian representation forms a more linear and robust latent manifold for color styles, enabling smooth manipulation between disparate color profiles.
It is worth noting that no additional constraints were applied to optimize blending during the training of all models. Thus, these results reflect the inherent blending capabilities of different LUT representations.

\begin{table}[ht]
\centering
\caption{Quantitative comparison of blending with different $\alpha$ on the 100 natural images from MIT5K~\cite{bychkovsky2011learning}.}
\label{tab:blending_7luts}
\resizebox{0.7\textwidth}{!}{%
\setlength{\tabcolsep}{4pt}
\begin{tabular}{c|l|cccccc}
\toprule
Metric & \multicolumn{1}{c|}{Method} & 0 & 0.2 & 0.4 & 0.6 & 0.8 & 1 \\ \midrule
\multirow{5}{*}{PSNR$\uparrow$} & CNILUT (128$\times$3) & 39.97 & 26.80 & 23.37 & 23.99 & 27.71 & 40.27 \\
 & CNILUT (256$\times$3) & 44.30 & 29.02 & 26.06 & 25.92 & 29.31 & 44.17 \\
 & ENNELUT ($L_2$) & \textit{48.29} & 33.50 & 30.29 & 30.62 & 34.35 & \textit{47.56} \\
 & CGLUT-32L (Full) & \textbf{48.67} & \textit{35.44} & \textit{31.16} & \textit{31.33} & \textit{34.64} & \textbf{47.95} \\
 & CGLUT-32L (Shared Geo.) & 47.36 & \textbf{38.46} & \textbf{34.67} & \textbf{34.47} & \textbf{37.60} & 46.18 \\ \midrule
\multirow{5}{*}{$\Delta E_{00}\downarrow$} & CNILUT (128$\times$3) & 1.87 & 7.88 & 10.70 & 10.13 & 7.13 & 1.15 \\
 & CNILUT (256$\times$3) & 1.21 & 7.27 & 9.72 & 9.69 & 7.11 & 0.80 \\
 & ENNELUT ($L_2$) & 0.81 & 3.61 & 5.14 & 4.89 & 3.14 & 0.77 \\
 & CGLUT-32L (Full) & \textbf{0.59} & \textit{3.08} & \textit{4.48} & \textit{4.70} & \textit{3.23} & \textbf{0.56} \\
 & CGLUT-32L (Shared Geo.) & \textit{0.66} & \textbf{2.11} & \textbf{3.19} & \textbf{2.96} & \textbf{1.94} & \textit{0.68} \\ \midrule
\multirow{5}{*}{$\Delta E_{76}\downarrow$} & CNILUT (128$\times$3) & 2.40 & 10.87 & 14.91 & 13.72 & 9.17 & 2.02 \\
 & CNILUT (256$\times$3) & 1.50 & 9.70 & 12.86 & 12.78 & 9.16 & 1.45 \\
 & ENNELUT ($L_2$) & 0.99 & 4.38 & 6.48 & 6.22 & 3.93 & 0.93 \\
 & CGLUT-32L (Full) & \textbf{0.79} & \textit{4.03} & \textit{5.86} & \textit{6.03} & \textit{4.06} & \textbf{0.73} \\
 & CGLUT-32L (Shared Geo.) & \textit{0.86} & \textbf{2.62} & \textbf{3.93} & \textbf{3.69} & \textbf{2.46} & \textit{0.87} \\ \midrule
\multirow{5}{*}{LPIPS$\downarrow$} & CNILUT (128$\times$3) & 0.015 & 0.104 & 0.166 & 0.154 & 0.098 & 0.010 \\
 & CNILUT (256$\times$3) & 0.008 & 0.086 & 0.130 & 0.131 & 0.086 & 0.007 \\
 & ENNELUT ($L_2$) & \textit{0.004} & 0.036 & 0.066 & 0.063 & \textit{0.035} & 0.012 \\
 & CGLUT-32L (Full) & \textbf{0.003} & \textit{0.026} & \textit{0.049} & \textit{0.054} & \textit{0.035} & \textit{0.008} \\
 & CGLUT-32L (Shared Geo.) & \textit{0.004} & \textbf{0.011} & \textbf{0.025} & \textbf{0.024} & \textbf{0.014} & \textbf{0.007} \\
 \bottomrule
\end{tabular}%
}
\end{table}

\begin{figure}[h]
    \centering
    \includegraphics[width=\linewidth]{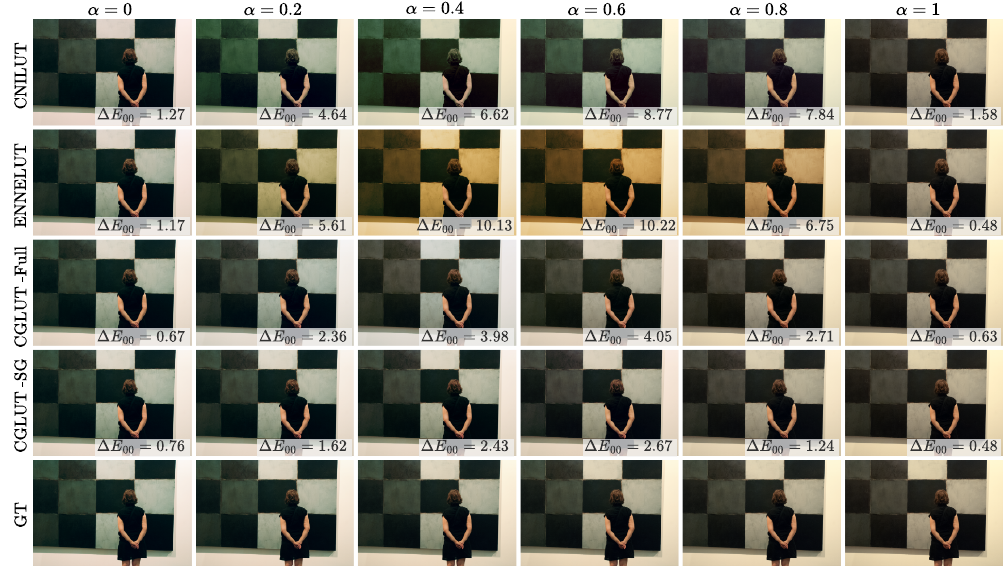}
    \bigskip
    \includegraphics[width=\linewidth]{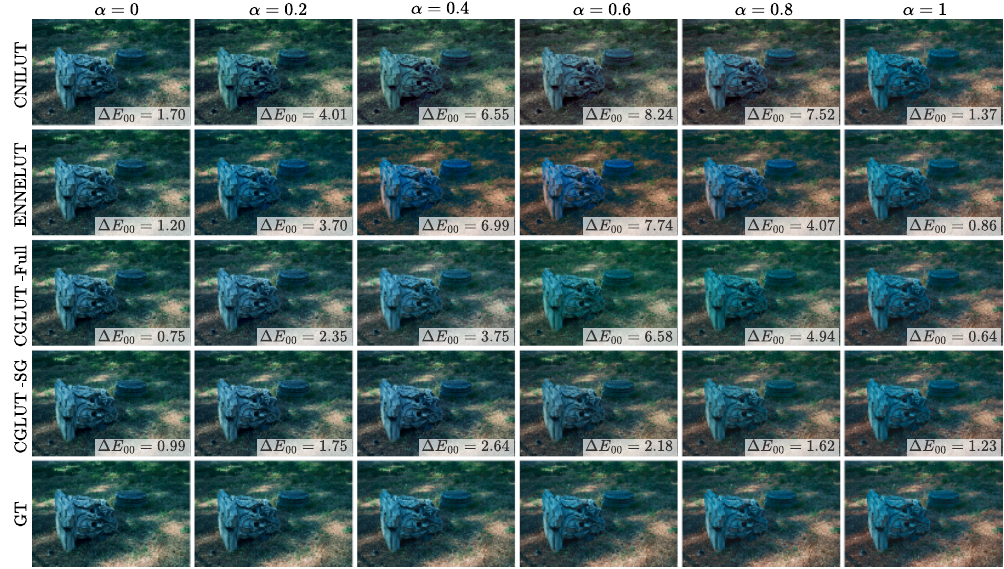}
    \caption{Qualitative comparison of style blending with different $\alpha$. Rows 1–5 illustrate the results of CNILUT (256$\times$3), ENNELUT ($L_2$), CGLUT-32 (Full generation), CGLUT-32 (Shared Geometry), and direct image-space blending (ground-truth), respectively.}
    \label{sup_fig:blend}
\end{figure}

\Cref{sup_fig:blend} compares the LUT blending results between CNILUT \cite{conde2024nilut}, ENNELUT \cite{zehtab2025efficient}, and two setups of our CGLUT (\textit{Full Generation} and \textit{Shared Geometry}). The visualization reveals that CNILUT tends to introduce undesired color shifts that deviate from the two anchor styles. In contrast, ENNELUT and CGLUT produce smoother transitions while better preserving the characteristics of the anchor styles.

Regarding implicit models, we hypothesize that ENNELUT achieves better blending results compared to CNILUT because it injects the LUT index into every residual block of the network. In contrast, CNILUT only provides the index to the initial layer, which limits the network's ability to leverage index information. Unlike these implicit approaches, CGLUT explicitly generates Gaussian parameters to construct each LUT. Interpolating the style embeddings therefore provides a direct mechanism for blending the corresponding LUT representations.

\begin{table}[h]
\caption{Quantitative comparison of CGLUT under \textit{Shared Geometry} ($\dagger$) and \textit{Full Generation}.}
\label{tab:cglut_sg}
\resizebox{\columnwidth}{!}{%
\setlength{\tabcolsep}{2pt}
\begin{tabular}{l|ccccc|ccccc|ccccc}
\toprule
\multicolumn{1}{c|}{\multirow{2}{*}{Method}} & \multicolumn{5}{c|}{7 LUTs}                                   & \multicolumn{5}{c|}{75 LUTs}                                  & \multicolumn{5}{c}{225 LUTs}                                 \\ 
\multicolumn{1}{c|}{}                        & PSNR$\uparrow$  & $\Delta E_{00}\downarrow$ & $\Delta E_{76}\downarrow$ & \#Params  & GFLOPs & PSNR$\uparrow$  & $\Delta E_{00}\downarrow$ & $\Delta E_{76}\downarrow$ & \#Params  & GFLOPs & PSNR$\uparrow$  & $\Delta E_{00}\downarrow$ & $\Delta E_{76}\downarrow$ & \#Params  & GFLOPs  \\ \midrule

CGLUT-32$\dagger$ (Small)                              & 49.35 & 0.25           & 0.49           & 58K  & 0.49   & 43.55 & 0.49           & 0.95           & 62K  & 0.49   & 41.97 & 0.69           & 1.38           & 72K  & 0.49   \\
CGLUT-64$\dagger$ (Small)                              & 52.02 & 0.19           & 0.36           & 85K  & 0.98   & 46.89 & 0.35           & 0.65           & 89K  & 0.98   & 44.81 & 0.49           & 0.99           & 99K  & 0.98  \\
CGLUT-32$\dagger$ (Large)                       & 49.58 & 0.24            & 0.48            & 163K & 0.49   & 43.79 & 0.48            & 0.93            & 168K & 0.49   & 44.02 & 0.56            & 1.12            & 177K & 0.49   \\
CGLUT-64$\dagger$ (Large)                        & 52.62 & 0.18            & 0.35            & 217K & 0.98   & 47.13 & 0.34            & 0.64            & 222K & 0.98   & 47.54 & 0.37            & 0.74            & 231K & 0.98   \\ \midrule
CGLUT-32 (Small)                             & 50.76 & 0.21           & 0.41           & 84K  & 0.49   & 45.66 & 0.39           & 0.73           & 89K & 0.49   & 46.68 & 0.41           & 0.82           & 98K  & 0.49   \\

CGLUT-64 (Small)                             & 54.06 & 0.16           & 0.32           & 130K & 0.98   & 48.43 & 0.30           & 0.55           & 135K & 0.98   & 49.39 & 0.30           & 0.59           & 144K & 0.98  \\ 
CGLUT-32 (Large)                            & 53.55 & 0.17            & 0.34            & 233K & 0.49   & 46.69 & 0.36            & 0.66            & 238K & 0.49   & 49.84 & 0.30            & 0.58            & 247K & 0.49   \\

CGLUT-64 (Large)                             & 55.10 & 0.14            & 0.29            & 324K & 0.98   & 49.37 & 0.28            & 0.51            & 328K & 0.98   & 52.41 & 0.22            & 0.43            & 338K & 0.98    \\ 

\bottomrule
\end{tabular}%
}
\end{table}

\Cref{tab:cglut_sg} compares two CGLUT configurations for multi-LUT fitting: \textit{Shared Geometry} and \textit{Full Generation}. While shared geometry achieves better blending quality with fewer parameters, its multi-LUT fitting accuracy is lower than that of the \textit{full generation} model. This is because sharing spatial attributes across all Gaussian primitives limits the model's expressiveness. Notably, this decline in LUT reconstruction quality becomes more pronounced as the number of LUTs increases. Therefore, for applications where blending is not required, we recommend the \textit{full generation} setting to leverage its more flexible Gaussian representations.


\subsection{LUT Editing Results}
\label{sup_sec:edit}

\subsubsection{Additional GLUT Editing Results}

\begin{figure}[ht]
    \includegraphics[width=\linewidth]{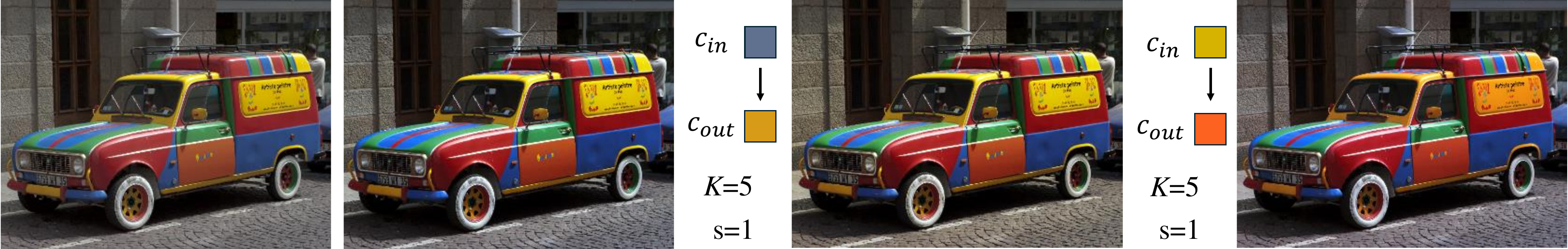}
    \includegraphics[width=\linewidth]{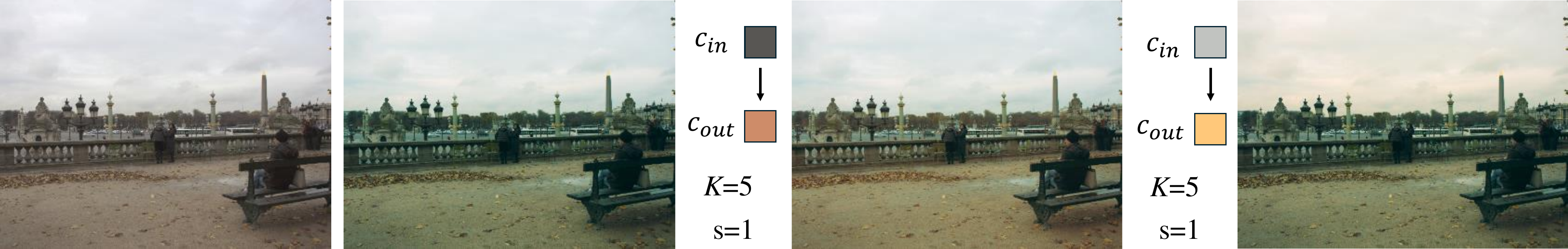}    
    \includegraphics[width=\linewidth]{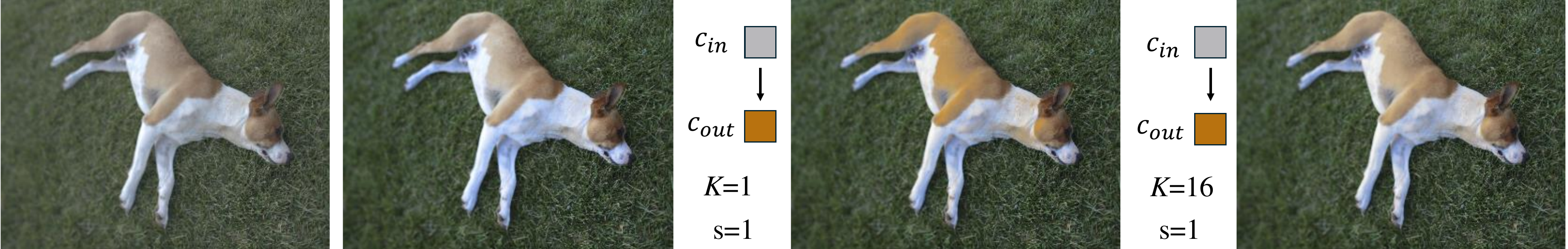}
    \includegraphics[width=\linewidth]{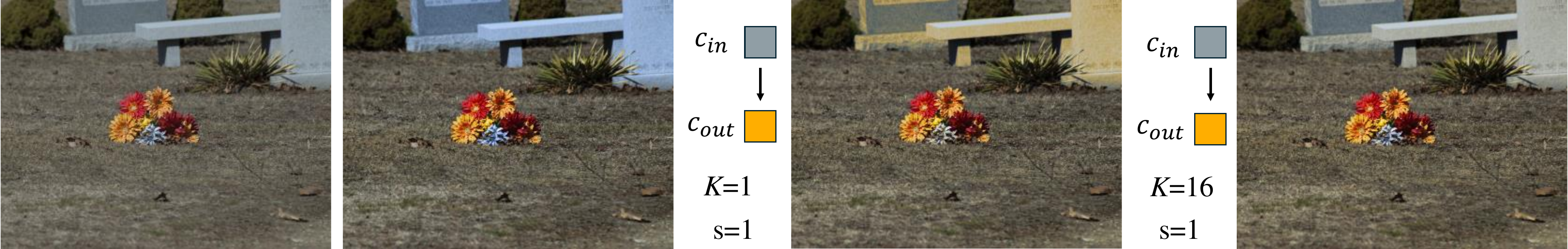}
    \includegraphics[width=\linewidth]{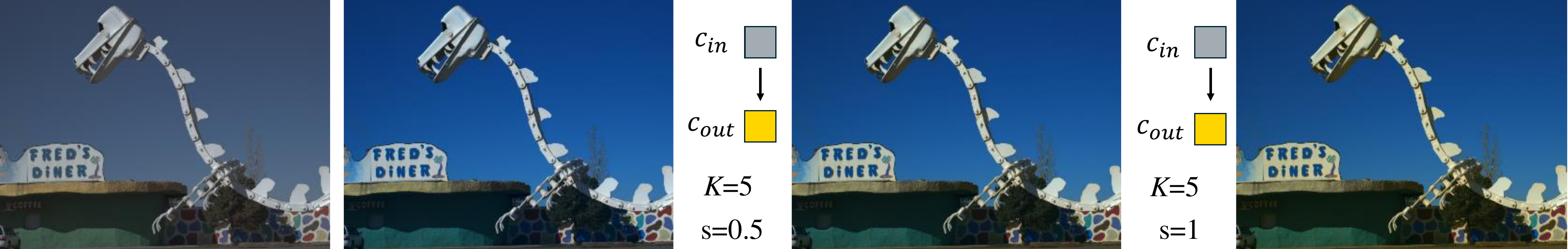}
    \includegraphics[width=\linewidth]{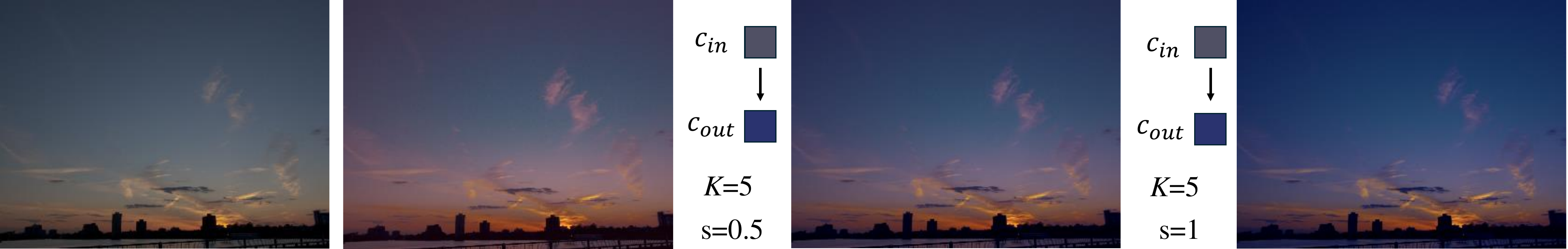}
    \makebox[0.21\textwidth]{\scriptsize Input}
    \makebox[0.21\textwidth]{\scriptsize GLUT}
    \makebox[0.07\textwidth]{\scriptsize }
    \makebox[0.21\textwidth]{\scriptsize Editing 1}
    \makebox[0.06\textwidth]{\scriptsize }
    \makebox[0.21\textwidth]{\scriptsize Editing 2}
    \caption{Qualitative results of editing GLUT representations. Each row shows an input image, a retouched image by GLUT, and two edited versions obtained with different color constraints $\mathbf{c} = (\mathbf{c}_\text{in}, \mathbf{c}_\text{out})$, number of Gaussians to edit $K$, and strength $s$.}
    \label{fig:editing}
\end{figure}

\Cref{fig:editing} presents additional editing results under various configurations. Specifically, we evaluate the impact of different color constraints (Rows 1–2), the number of edited primitives $K$ (Rows 3–4), and the manipulation strength $s$ (Rows 5–6) on a single input image. The results demonstrate that our method consistently produces natural-looking color transitions. Since $K$ determines the number of primitives involved in the update, a larger $K$ spreads the edit across a broader region of the color manifold, thereby affecting a wider range of colors while minimizing abrupt shifts in specific hues. Similarly, the strength $s$ controls the degree to which the source color $\mathbf{c}_\text{in}$ is displaced toward the target $\mathbf{c}_\text{out}$. A higher $s$ yields a result closer to the target color. Furthermore, applying edits iteratively on the output image allows for a progressively more pronounced and customized effect.

\subsubsection{Locality evaluation}
\label{sup_sec:locality}

To evaluate locality, we compute the color change before and after each edit at every node of a regular $\text{33}^\text{3}$ RGB grid, measured as $\Delta E_{00}\big(f_\text{before}(\mathbf{x}), f_\text{after}(\mathbf{x})\big)$. We pool the nodes across all edits and divide them into 20 bins according to their Euclidean distance $\|\mathbf{x}-\mathbf{c}_\text{in}\|_2$, which ranges from 0 to $\sqrt{3}\approx1.73$. For each bin, we report the mean color change. We define the leakage radius as the center of the first bin at which the mean color change falls below one JND.

\textbf{Comparison with fine-tuning an implicit LUT.}
We compare GLUT editing with fine-tuning NILUT~\cite{conde2024nilut} under identical editing constraints. For each of the seven LUTs, we use a GLUT-32 and a NILUT (128$\times$2) trained on the same Hald image pair. For each LUT, we sample 50 source colors $\mathbf{c}_\text{in}$ uniformly from the RGB cube. We obtain each target color $\mathbf{c}_\text{out}$ by perturbing the reference LUT output at $\mathbf{c}_\text{in}$ by 10--20 $\Delta E_{00}$ in a random direction in CIELAB, simulating a local color adjustment. Both models are thus evaluated on the same set of 350 edits.

For GLUT, we apply each edit with $K=1$ and $s=1$. A single application may move $f(\mathbf{c}_\text{in})$ only partway toward $\mathbf{c}_\text{out}$, so we repeat the edit until the $\Delta E_{00}$ between $f(\mathbf{c}_\text{in})$ and $\mathbf{c}_\text{out}$ falls below one JND. This requires an average of about 10 repetitions and at most 30.

For NILUT, we fine-tune all network weights $m$ using Adam with a learning rate of $5\times10^{-4}$ and the following loss:
\begin{equation}
\mathcal{L} = \mathbb{E}_{\mathbf{x}\sim\mathcal{U}([0,1]^3)}\,\|m(\mathbf{x}) - m_0(\mathbf{x})\|_1 + \lambda\,\|m(\mathbf{c}_\text{in}) - \mathbf{c}_\text{out}\|_1,
\qquad \lambda = 10,
\end{equation}
where $m_0$ is a frozen copy of the pretrained network. The first term encourages the original mapping to be preserved and is estimated using 4096 newly sampled uniform RGB values at each step. The second term encourages the output at $\mathbf{c}_\text{in}$ to match the target. We stop fine-tuning when the $\Delta E_{00}$ at $\mathbf{c}_\text{in}$ falls below 1.0 or after 8000 steps. Within this budget, NILUT fails to reach the target for 35 of the 350 edits, whereas GLUT reaches it for all edits. We exclude these 35 edits from both methods so that the locality analysis compares only edits successfully completed by both.

\begin{figure}[h]
    \centering
    \includegraphics[width=0.45\linewidth]{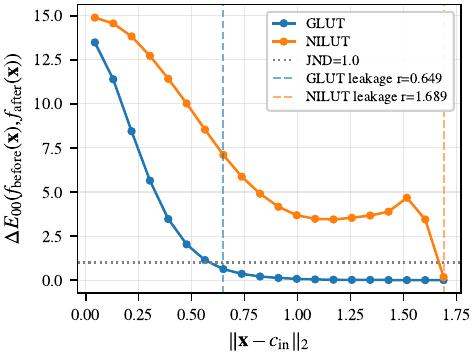}
    \caption{Edit locality of GLUT and NILUT. Mean $\Delta E_{00}$ change as a function of the RGB distance to the source color $\mathbf{c}_\text{in}$, pooled over 315 edits on 7 LUTs. The dashed line marks one JND.}
    \label{fig:locality}
\end{figure}

\Cref{fig:locality} shows the resulting curves for GLUT and NILUT. For both methods, the mean color change decreases with distance from $\mathbf{c}_\text{in}$, but it decays much faster for GLUT. The leakage radius is 0.65 for GLUT and 1.69 for NILUT. For NILUT, the mean change remains above 10 $\Delta E_{00}$ across most of the RGB cube, whereas for GLUT it approaches zero at a distance of approximately 1.0. This difference is consistent with NILUT updating over 80\% of its network weights, indicating that the edit is not localized in parameter space and offers no explicit control over its spatial extent in RGB space.

This difference arises even though NILUT is fine-tuned with an explicit regularization term to preserve the original mapping, whereas GLUT editing uses no such term. The difference stems from the two representations. A GLUT edit modifies only the biases of the $K$ primitives with the largest weights at $\mathbf{c}_\text{in}$, so its effect is governed by their influence over RGB space. NILUT lacks this explicit local structure, and fine-tuning its fully connected weights can alter the mapping far from $\mathbf{c}_\text{in}$.

\begin{figure}[h]
    \centering
    \begin{subfigure}[b]{0.45\linewidth}
        \centering
        \includegraphics[width=\linewidth]{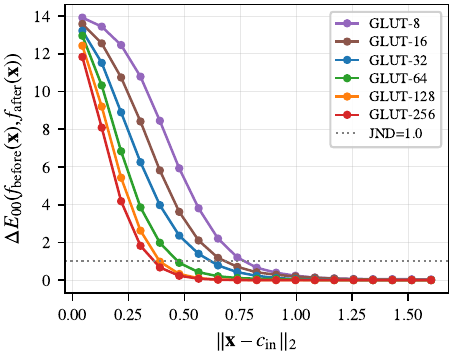}
        \caption{Number of Gaussians $N$}
        \label{fig:locality_N}
    \end{subfigure}
    \hfill
    \begin{subfigure}[b]{0.46\linewidth}
        \centering
        \includegraphics[width=\linewidth]{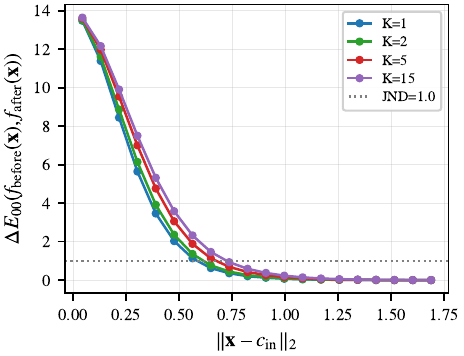}
        \caption{Number of edited primitives $K$}
        \label{fig:locality_K}
    \end{subfigure}
    \caption{Edit locality of GLUT. (a) GLUTs with different numbers of Gaussians fitted to the same LUT, edited with $K=1$. (b) A single GLUT edited with different $K$.}
    \label{fig:locality_param}
\end{figure}

\textbf{Influence of $N$ and $K$ on Editing Locality.} \Cref{fig:locality_param} shows how the locality of GLUT editing depends on the number of Gaussians $N$ and the number of edited primitives $K$. In \Cref{fig:locality_N}, we fit GLUTs with different values of $N$ to the same LUT and edit them with $K=1$. As $N$ increases, the curves decay faster, and the leakage radius decreases. This is because, with more Gaussians, each primitive generally covers a smaller region of the RGB cube, causing its influence to decay over a shorter distance and keeping the edit more localized. In \Cref{fig:locality_K}, we edit a single GLUT with different values of $K$. Smaller values of $K$ produce steeper decay, although the curves remain close across different values of $K$. This is because the additional primitives selected at larger $K$ have relatively small weights at $\mathbf{c}_\text{in}$ and therefore receive only a small share of the edit, resulting in only a modest change in its spatial extent. 

\begin{figure}[h]
    \centering
    \includegraphics[width=\linewidth]{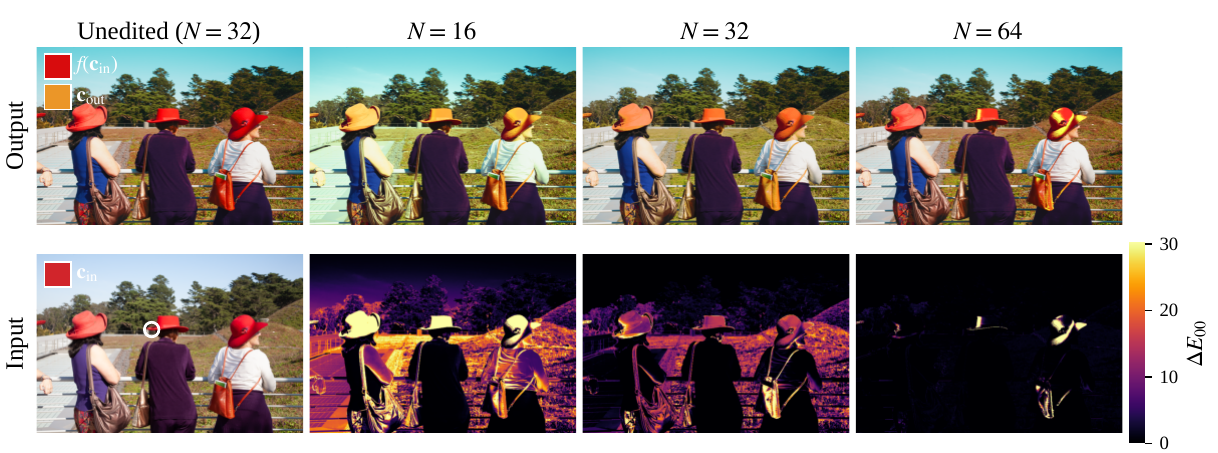}
    \caption{An editing example with different numbers of Gaussians $N$.}
    \label{fig:edit_n}
\end{figure}

\begin{figure}[h]
    \centering
    \includegraphics[width=\linewidth]{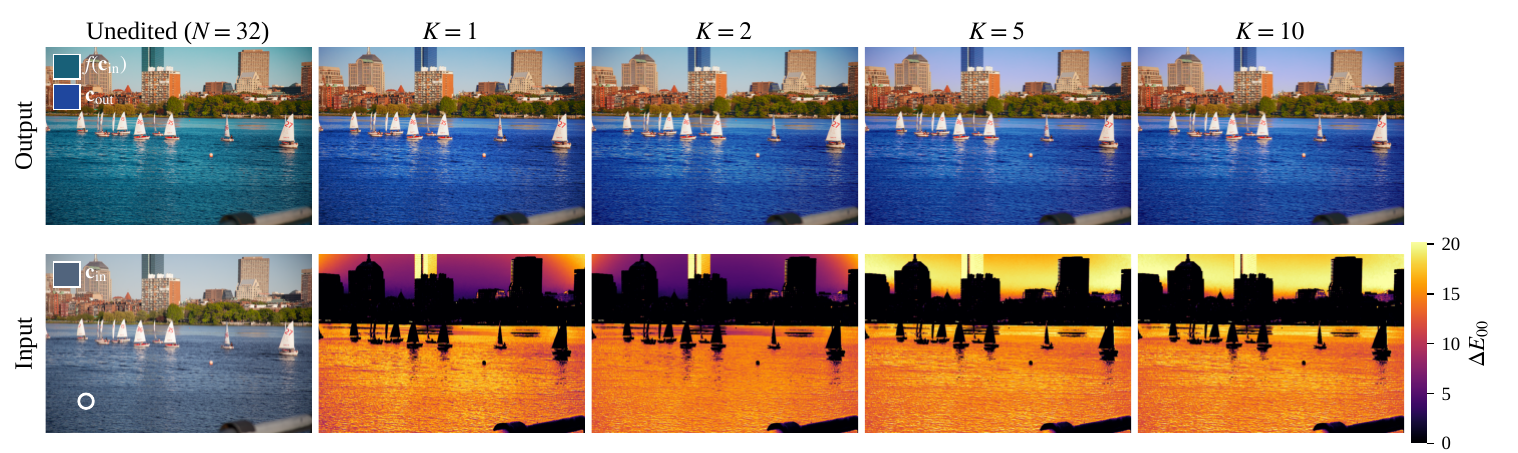}
    \caption{An editing example with different numbers of edited primitives $K$.}
    \label{fig:edit_k}
\end{figure}

\Cref{fig:edit_n} and \Cref{fig:edit_k} illustrate these trends for a single edit. We apply the edit with $s=1$ and repeat it until $\Delta E_{00}$ between the current output at $f(\mathbf{c}_\text{in})$ and the target $\mathbf{c}_\text{out}$ falls below one JND, ensuring that all settings reach the same target. This experiment illustrates the effect of $N$ and $K$ under a common stopping criterion. In typical editing scenarios, users may stop after fewer iterations once the desired result is achieved. The top row shows the edited outputs, and the bottom row shows the corresponding $\Delta E_{00}$ between the edited and unedited outputs.

\Cref{fig:edit_n} varies $N$ with $K=1$. We select a red color from the hat and turn it orange. We fit GLUT with $N=16$, $32$, and $64$ to the same style and apply each model to the input image. As $N$ increases, each primitive generally covers a smaller region of RGB space, making the edit more localized. With GLUT-16, the edit also changes the sky, the white shirt, and the grass, whereas with GLUT-64, it is largely confined to the hats. However, this increased locality comes at a cost: the shaded parts of the hats retain their original red, while other parts turn yellow. Since GLUT operates solely on RGB values without accounting for semantic regions, it cannot ensure consistent color changes across the same object. Incorporating semantic constraints into the editing process is a promising direction for future work.

\Cref{fig:edit_k} varies $K$ with GLUT-32. We select a green-blue color from the sea and shift it toward ocean blue. As $K$ increases, the edit affects more regions, including the sky. The results for $K=5$ and $K=10$ are almost identical because the additional primitives have small weights at $\mathbf{c}_\text{in}$ and therefore receive only a small share of the edit. Both observations are consistent with the locality curves in \Cref{fig:locality_K}.

\textbf{Locality on Overlapping Gaussians.}
We investigate whether editing becomes less effective in regions where multiple Gaussians have comparable influence. We define the ambiguity of a color $\mathbf{x}$ as $a(\mathbf{x}) = w_{(1)}(\mathbf{x}) / w_{(2)}(\mathbf{x})$, the ratio of its largest to second-largest weight. Values close to one indicate substantial overlap, with multiple primitives contributing similarly, whereas large values indicate that a single primitive dominates. Using GLUT-32 trained on the 75-LUT dataset, we compute $a(\mathbf{x})$ on a $64^3$ grid and sample 50 overlap points with $a(\mathbf{x}) \le 1.5$ and 50 dominated points with $a(\mathbf{x}) \ge 10$. For each point, we specify a target $\mathbf{c}_\text{out}$ whose $\Delta E_{00}$ distance from $f(\mathbf{c}_\text{in})$ is between 10 and 20. We then apply edits with $K=1$ and $s=1$, repeating each edit until the error falls below one JND or 30 iterations are reached.

All 100 edits reach the error threshold within 30 iterations, indicating that overlap does not prevent successful editing under our experimental settings. However, overlap slows convergence: overlap points require an average of 13.3 iterations, compared with 3.4 for dominated points. When the weights at $\mathbf{c}_\text{in}$ are more evenly distributed, each edit corrects a smaller fraction of the residual. Overlap also reduces locality. The mean leakage radius is 0.776 for overlap points and 0.584 for dominated points. This is consistent with the edit affecting multiple primitives, potentially extending its influence across the regions they cover.

\subsection{Ablation Studies}
\label{sup_sec:ablation}

We conduct several ablation studies to verify the proposed Gaussian-based LUT representation. We train the GLUT-32 model under different settings and test it on the 75-LUT Hald image dataset.

\subsubsection{Number of Gaussian Primitives}

\begin{figure}[ht]
    \centering
    \begin{minipage}[c]{0.58\textwidth}
        \centering
        \captionof{table}{Ablations on the number of Gaussian primitives.}
        \label{tab:ablation_num}
        \resizebox{0.8\textwidth}{!}{%
        \begin{tabular}{ccccc}
        \toprule
        \#Gaussian  & PSNR$\uparrow$   & $\Delta E_{00}\downarrow$  & $\Delta E_{76}\downarrow$ & \#Params\\ \midrule
         8  & 37.01 &1.068 &2.119 & 188 \\
         16  & 41.50	&0.636	&1.223  & 364\\ 
         32  & 45.47 & 0.414 & 0.770  & 716\\
         64  & 48.42 & 0.310 & 0.560 & 1,420\\
         128  & 50.31 & 0.263 & 0.464 & 2,828 \\
         \bottomrule
        \end{tabular}
        }
    \end{minipage}
    \hfill
    \begin{minipage}[c]{0.40\textwidth}
        \centering
        \includegraphics[width=0.9\textwidth]{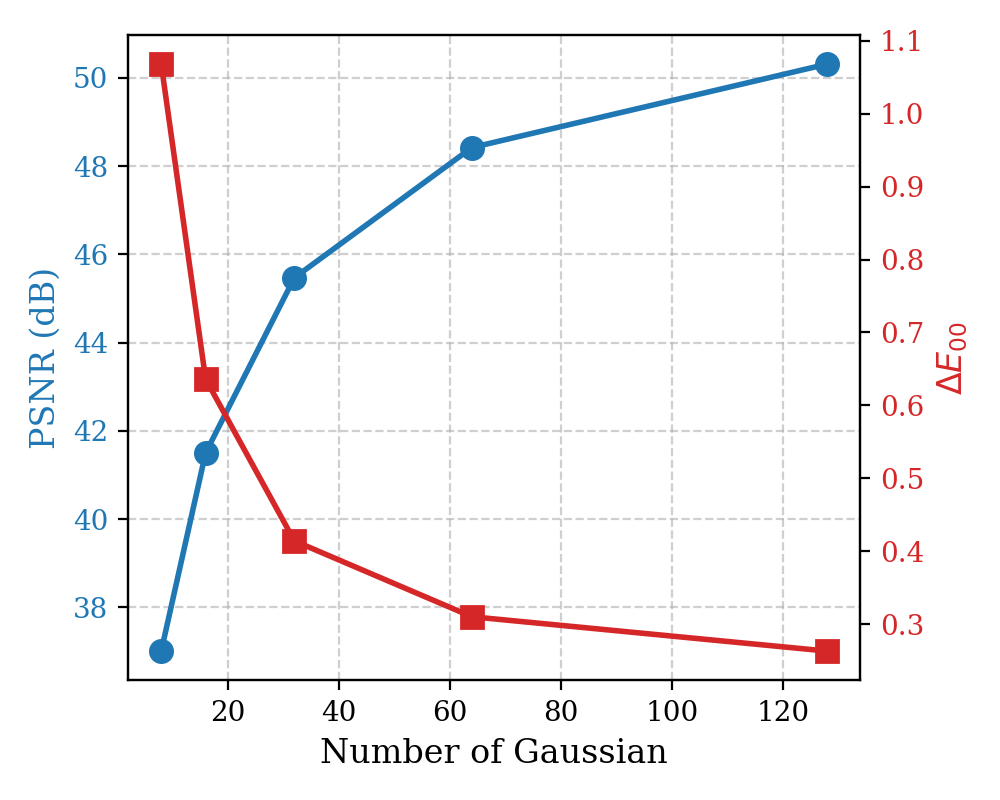}
        \captionof{figure}{Quantitative results of GLUT with different numbers of Gaussians.}
        \label{fig:ablation_num}
    \end{minipage}
\end{figure}

\Cref{tab:ablation_num} and \Cref{fig:ablation_num} present the performance of our GLUT across varying numbers of Gaussian primitives. As shown, the parameter size is linearly proportional to the number of Gaussians. While increasing the primitive count consistently improves performance, the growth rate of PSNR and the reduction in $\Delta E_{00}$ exhibit diminishing returns, suggesting an optimal balance between model complexity and reconstruction accuracy.

\subsubsection{Local Color Transformation}
\Cref{tab:ablation_local} shows different formulations for the local color mapping function, including a learnable RGB value, an affine transformation (our default), and second- and third-order Spherical Harmonics (SH) (with $9\times3=27$ and $16\times3=48$ parameters). These variants introduce differences in model capacity and color accuracy. Generally, we observe that performance is positively correlated with parameter size. Despite the higher representational power of SH bases, we select the affine transformation as our final design, as it shows a better tradeoff between color fidelity and efficiency, and also provides a more robust and intuitive representation for our GLUT editing in \Cref{sec:edit}.

\subsubsection{Other Components}
\Cref{tab:ablation_params} summarizes ablations on other key parameters. To verify the impact of weight normalization, we replace the weights in \Cref{eq:weights} by $w_i(\mathbf{x})=p_i(\mathbf{x})\, o_i$. The results in the first row show a decrease in PSNR and larger color differences when normalization is removed. To investigate the impact of opacity, we consider a configuration excluding $o_i$, where the weighted activation in \Cref{eq:weights} is replaced by a standard softmax-like normalization: $w_i(\mathbf{x}) = p_i(\mathbf{x}) / (\sum_{j=1}^{N} p_j(\mathbf{x}) + \epsilon)$. Moreover, we evaluate the role of the global transformation branch by comparing two variants: a residual connection without global transform, defined as $f(\mathbf{x}) = \sum_{i=1}^{N} w_i(\mathbf{x})f_i(\mathbf{x}) + \mathbf{x}$, and a pure local mapping that relies solely on weighted local transformations, $f(\mathbf{x}) = \sum_{i=1}^{N} w_i(\mathbf{x})f_i(\mathbf{x})$, compared to \Cref{eq:mapping}. 
The results underscore the effectiveness of residual learning. Specifically, predicting the color displacement relative to the input is significantly more efficient than directly regressing the absolute target color values. This allows the Gaussian primitives to focus on modeling the non-linear residuals. Furthermore, the learned opacity and global transformation also enhance the representation, resulting in higher color mapping fidelity.

\begin{table}[t]
    \centering
\caption{Ablation on Gaussian parameters.}
\begin{subtable}[b]{0.5\textwidth}
    \centering
    \resizebox{0.92\textwidth}{!}{%
    \begin{tabular}{lccccc}
    \toprule
    Function  & PSNR$\uparrow$  & $\Delta E_{00}\downarrow$ & $\Delta E_{76}\downarrow$ & \#Params  & GFLOPs\\ \midrule
    RGB  & 42.58 & 0.565 & 1.059 & 428 & 0.30 \\
    SH-2 & 45.67 & 0.405 & 0.740 & 1,196 & 0.73 \\
    SH-3 & \textbf{46.78} & \textbf{0.360} & \textbf{0.651} & 1,868 & 1.12  \\ 
    \textbf{Affine} & 45.47 & 0.414 & 0.770 & 716 & 0.49 \\
     \bottomrule
    \end{tabular}}
    \caption{Local color transformation functions.}
    \label{tab:ablation_local}
\end{subtable}%
\hfill
\begin{subtable}[b]{0.5\textwidth}
    \centering
    \resizebox{0.98\textwidth}{!}{%
    \begin{tabular}{lcccc}
    \toprule
    Configuration & PSNR $\uparrow$ & $\Delta E_{00} \downarrow$ & $\Delta E_{76} \downarrow$ & \#Params \\ \midrule
    w/o weight normalization & 45.17 &0.419 & 0.776 & 716 \\
    w/o Opacity ($o_i$)           & 45.43 & 0.415 & 0.774 & 684 \\
    w/o Global  ($\mathbf{G}, \mathbf{g}$)  & 45.28 & 0.419 & 0.776 & 704 \\
    w/o Global w/o Residual        & 40.54 & 0.661 & 1.273 & 704 \\ 
    \textbf{Full Model} & \textbf{45.47} & \textbf{0.414} & \textbf{0.770} & 716 \\
    \bottomrule
    \end{tabular}}
    \caption{Opacity and Global transformation.}
    \label{tab:ablation_params}
\end{subtable}%
\label{tab:ablation_param}
\end{table}

\begin{table}[t]
    \centering
\caption{Ablation on training settings.}
\begin{subtable}[b]{0.36\textwidth}
    \centering
    \resizebox{\textwidth}{!}{%
    \begin{tabular}{cccccc}
    \toprule
    $\mathcal{L}_{\text{rec}}$ & $\mathcal{L}_{\text{hc}}$ & $\mathcal{R}_{\text{sparse}}$   & PSNR$\uparrow$   & $\Delta E_{00}\downarrow$  & $\Delta E_{76}\downarrow$\\ \midrule
     \checkmark &     &          &45.36 & 0.425 & 0.785 \\
     \checkmark & \checkmark  &  & 45.38 & 0.419 & 0.776  \\
     \checkmark & \checkmark  & \checkmark  & \textbf{45.47} & \textbf{0.414} & \textbf{0.770}  \\ 
     \bottomrule
    \end{tabular}}
   \caption{Loss functions.}
    \label{tab:ablation_loss}
\end{subtable}%
\begin{subtable}[b]{0.32\textwidth}
    \centering
    \resizebox{0.9\textwidth}{!}{%
    \begin{tabular}{ccccc}
    \toprule
    Mining  & PSNR$\uparrow$   & $\Delta E_{00}\downarrow$  & $\Delta E_{76}\downarrow$\\ \midrule
       & 45.13 &0.422 &0.793  \\
     \checkmark  & \textbf{45.47} & \textbf{0.414} & \textbf{0.770}  \\ 
     \bottomrule
    \end{tabular}}
    \caption{Hard sample mining strategy.}
    \label{tab:ablation_mine}
\end{subtable}%
\begin{subtable}[b]{0.32\textwidth}
    \centering
    \resizebox{0.96\textwidth}{!}{%
    \begin{tabular}{cccc}
    \toprule
    Mean init  & PSNR$\uparrow$   & $\Delta E_{00}\downarrow$  & $\Delta E_{76}\downarrow$\\ \midrule
     Random  &45.45 &0.420 &0.782  \\
     Uniform  & \textbf{45.47} & \textbf{0.414} & \textbf{0.770}  \\ 
     \bottomrule
    \end{tabular}}
    \caption{Mean value initialization.}
    \label{tab:ablation_init}
\end{subtable}%
\label{tab:ablation_train}
\end{table}

\subsubsection{Loss Terms}
\Cref{tab:ablation_loss} shows the contribution of each loss component. Starting with the basic reconstruction loss $\mathcal{L}_{\text{rec}}$, we observe that the introduction of the hue-chroma loss $\mathcal{L}_{\text{hc}}$ effectively reduces color distortion, leading to lower $\Delta E_{00}$ and $\Delta E_{76}$ metrics. Furthermore, the inclusion of the sparsity regularization $\mathcal{R}_{\text{sparse}}$ yields the best overall performance on both PSNR and $\Delta E$.

\subsubsection{Hard Sample Mining}

To investigate the effect of hard sample mining, we compare the baseline model with and without this strategy. As shown in \Cref{tab:ablation_mine}, applying hard sample mining brings consistent improvements across all metrics. These results demonstrate that focusing on challenging samples during training effectively guides the model to achieve better color fidelity.

\subsubsection{Initialization}

\Cref{tab:ablation_init} compares the impact of random and uniform initialization on the Mean parameter. Since uniform initialization performs marginally better, we adopt it as our default configuration. Notably, these results also demonstrate that our GLUT can consistently converge to an optimal distribution of Gaussians regardless of the specific initialization strategy, highlighting the robustness of our optimization process.

\section{Interactive GLUT-based LUT editing tool}
\label{sup_sec:editing}

\begin{figure}[ht]
    \centering
    \includegraphics[width=\linewidth]{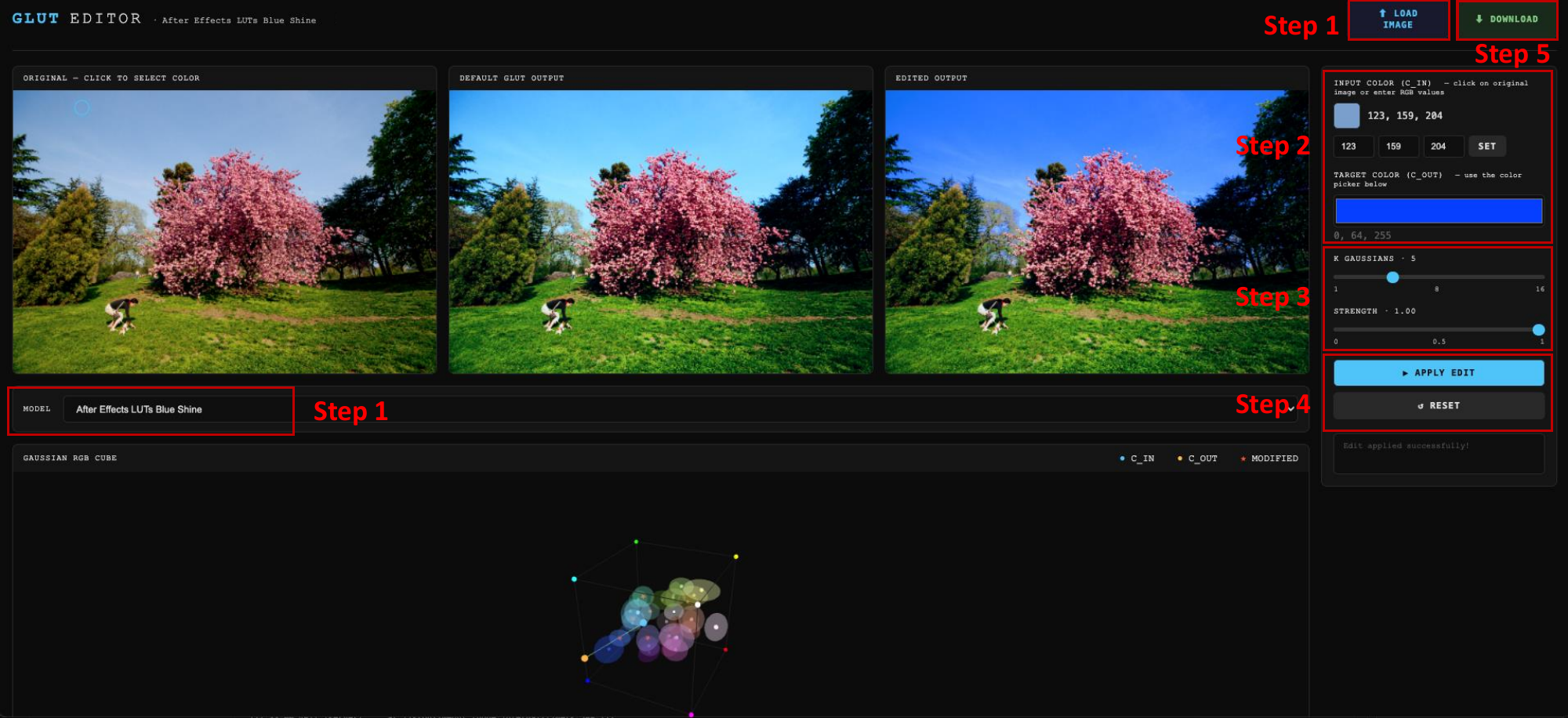}
    \caption{Snapshot of the interactive GLUT editing tool. Red numbers correspond to the operation steps.}
    \label{fig:gui}
\end{figure}

In the supplementary material, we provide an interactive demo for the LUT editing tool by implementing the GLUT editing method described in \Cref{sec:edit} of the main paper. The demo runs entirely in the browser (HTML/JavaScript, CPU-only, no external dependencies) in real time.
\Cref{fig:gui} shows a snapshot of the interactive LUT editing tool. A step-by-step guide is provided below, with each step referenced by the corresponding red number annotated in \Cref{fig:gui}.

The operation instructions of the tool are as follows:

\textbf{Step 1: Load Image and GLUT Model.}
Load a source image by clicking the blue ``LOAD IMAGE'' button in the top-right corner. Select a pretrained GLUT model from the drop-down menu. The interface will display the original input, the retouched output, and the edited output in the ``ORIGINAL'', ``DEFAULT GLUT OUTPUT'', and ``EDITED OUTPUT'' panels, respectively. Note that by default the ``EDITED OUTPUT'' image is the same as the ``DEFAULT GLUT OUTPUT''. The selected model name (\eg, ``\textit{After Effects LUTs Blue Shine}'') will appear at the top. If no model is active, it displays ``\textit{No model loaded}''. The ``GAUSSIAN RGB CUBE'' panel visualizes each Gaussian primitive using its learned mean RGB value. Primitives with means outside the standard RGB unit cube are visualized in gray.

\textbf{Step 2: Define Source ($\mathbf{c}_\text{in}$) and Target ($\mathbf{c}_\text{out}$) Colors.}
There are two ways to specify the source color $\mathbf{c}_\text{in}$: (i) Directly click on the original image to select the pixel color, or (ii) Manually enter RGB values in the right-hand control panel. To define the target color $\mathbf{c}_\text{out}$, click the ``TARGET COLOR'' box and use the color picker. The selection is visualized in the RGB cube as a blue node ($\mathbf{c}_\text{in}$) and an orange node ($\mathbf{c}_\text{out}$), connected by a vector illustrating the transformation.

\textbf{Step 3: Configure Parameters.}
Adjust the number of affected Gaussians ($K$) and the manipulation strength using the sliders on the right-hand panel. The parameter $K$ controls the spatial extent of the edit: a larger $K$ involves more Gaussian primitives in the update, thereby affecting a broader region of the color manifold. The Strength slider modulates the intensity of the transformation. Increasing this value pushes the source color $\mathbf{c}_\text{in}$ closer to the specified target $\mathbf{c}_\text{out}$, allowing for both subtle refinements and bold artistic changes.

\textbf{Step 4: Apply and Reset Edits.}
Click ``APPLY EDIT'' to perform the localized transformation. The result will be displayed in the ``EDITED OUTPUT'' panel. At the same time, the centers of the modified Gaussian primitives will be highlighted as red stars in the ``GAUSSIAN RGB CUBE'' panel. Subsequent clicks will apply the edit incrementally. To revert all changes and return to the original GLUT-retouched version, press the ``RESET'' button.

\textbf{Step 5: Export Results.}
Click the ``DOWNLOAD'' button in the top-right corner to save the edited image to the local device.

